%% file: main.tex
\documentclass{article}
\usepackage[final,nonatbib]{cpal_2025}
\usepackage[authoryear]{natbib}
\usepackage{tikz}
\usetikzlibrary{arrows.meta,positioning,fit,calc}
\usepackage{graphicx}      % for \resizebox
\usepackage{subcaption}
\usepackage{amssymb}
\usepackage{tabularx}
\usepackage{booktabs}
\usepackage{multirow}
\usepackage{float}
\usepackage{makecell}
\usepackage[table]{xcolor}

\input{math_commands.tex}

\usepackage{hyperref}
\usepackage{url}

\title{Absorbing State Phase Transitions \\in Multi-Agent Search}

\author{
Wenwen Zheng\thanks{Equal contribution.}\\
University of California, Santa Barbara\\
\texttt{wenwenzheng@ucsb.edu}\\
\And
Yuzhe Yang\footnotemark[1]\\
University of California, Santa Barbara\\
\texttt{yuzheyang@ucsb.edu}\\
\And
Helen Qu\footnotemark[1]\\
Flatiron Institute\\
\texttt{hqu@flatironinstitute.org}\\
\AND
Xin Eric Wang\\
University of California, Santa Barbara\\
\texttt{ericxwang@ucsb.edu}\\
\And
Haewon Jeong\\
University of California, Santa Barbara\\
Flatiron Institute\\
\texttt{haewon@ucsb.edu}\\
}
\newcommand{\ci}[3]{$#1^{+#2}_{-#3}$}

\begin{document}

\maketitle

\begin{abstract}
Nontrivial dynamics can emerge in large language model (LLM)-based multi-agent systems, and preliminary evidence exists that formalisms from statistical mechanics can be effective at modeling and predicting such behaviors.
In parallel, designing multi-agent communication topology for optimal task-solving is an active research question.
In this paper, we focus on predicting the success of multi-agent search tasks using the formalism of absorbing state phase transitions.
We first taxonomize search tasks into four types, informed by classical results in combinatorial search.
We then theoretically derive a critical communication degree \(d_c\), the minimum number of agents each agent can communicate with, above which incorrect hypotheses do not proliferate uncontrollably and the search enters the solved state.
Finally, we evaluate frontier LLM-based multi-agent systems on real-world search and discovery tasks, software configuration debugging and physical mechanism discovery, and find that agreement with theory is mixed.
LLM agents may not communicate with their neighbors and can develop strategies that are individually beneficial but limits the benefits of collaboration.
\footnote{Code: \url{https://github.com/wenwenzheng-gif/absorbing-state-multi-agent-search}}
\end{abstract}

\section{Introduction}

AI agents are increasingly being deployed in multi-agent systems, in which agents communicate, interact, and cooperate to complete tasks.
Multi-agent systems have been shown to improve over single agents across a variety of domains and task types \citep{wu2023autogen,guo2024multiagents,hong2024metagpt,qian2024chatdev,du2024multiagent,kim2024mdagents,agashe2025llm}.
However, comparatively little is known about their collective dynamics, such as emergent or nonlinear behavior; and effective modeling of these phenomena could improve the system's capability and safety.
While recent work suggests that formalisms like the Ising model from statistical mechanics \citep[e.g.][]{Ising1925} can be effective at modeling nonlinear multi-agent system dynamics \citep{el2026physicsagentsstatisticalmechanics,niu2026reliability,xie2026errorcascades,tanaka2026collectiveintelligencelotterymultiagent,pavlova2026flaggametoymodel}, these works have focused only on consensus formation and error propagation.

Beyond opinion dynamics, search problems provide another important setting for understanding collective behavior in multi-agent systems. 
In his famous Bitter Lesson post, Sutton wrote that ``\textit{Search and learning are the two most important classes of techniques for utilizing massive amounts of computation in AI research}" \citep{sutton2019bitter}.
Many tasks, from program synthesis \citep{gulwani2017program} to scientific discovery \citep{morales2026nmr}, involve exploring a space of candidate solutions through successive refinement and verification. 
In cooperative search, agents can explore different regions simultaneously and share findings that guide one another's exploration. 
Understanding these dynamics could help predict when collaboration accelerates discovery and when it leads to redundant effort or the propagation of unproductive search directions.

A crucial question in the design of multi-agent systems is inter-agent communication topology design \citep{zhang2024gdesigner,zhu2025multiagentbench,kim2025scaling}.
A fully dense communication graph may seem like a clear choice, but this is often prohibitively computationally expensive and can underperform sparser topologies \citep{li2024improving}.
Recent work has also shown that the optimal topology depends on the type of task the agents are asked to solve \citep{rizvimartel2025benefits,kim2025scaling,tang2025taskcomplexity,qian2025macnet,yang2025topological}.
Inter-agent communication should be most beneficial to solving search problems when evidence produced by one agent can affect the search space of others; however, there is little work demonstrating the effect of different inter-agent communication regimes on different search task types.

% There is also substantial theoretical work connecting search problems to absorbing state phase transitions \citep{semerjian2003randomwalksat,lee2010finitesize}
% \todo{more detwail}.

% In this paper, we investigate whether absorbing-state search theory can model LLM-based multi-agent search and predict the critical communication degree above which the population of incorrect candidates decays, allowing search to reach a solved state.
In this paper, we investigate multi-agent search dynamics through absorbing state phase transition theory, where search success guarantee is the absorbing state.
We introduce a taxonomy for search problems informed by classical results in parallel search and constraint satisfaction \citep{dechter1986learning,lai1984anomalies,dechter2007and}, distinguishing four task types: needle, chain, hierarchical, and factorizable search (Section~\ref{sec:taxonomy}).
We derive theoretical values for the critical communication degree $d_c$ for each of these task types (Section~\ref{sec:theory}), finding that factorizable search benefits the most from inter-agent communication due to the potential for information reuse across disparate parts of the search graph.
Finally, we evaluate on three factorizable search tasks using deterministic agents as well as LLM-based agents (Section~\ref{sec:exps}).
We find that deterministic, rule-based agent behaviors agree with theory, but current LLMs exhibit behaviors under which theoretical assumptions break down, such as individual strategizing to choose the best candidates, herding behavior amongst agents, and failure to communicate with their neighbors (Section~\ref{sec:results}).

\section{Setup}
\label{sec:setup}

\input{fig1}

\subsection{Problem Setup: Search Tasks}
We consider multi-agent systems applied to solving search tasks.
We formalize search tasks as a graph $\G(\cands, \mathcal{E})$ consisting of nodes $\cands$, representing the set of search candidates, and a set of directed edges $\mathcal{E}$ indicating dependency between candidates. We assume the graph has branching factor $b$. There is a single correct answer $c^{\star}$, the target of the search. We assume there exists a verification mechanism (e.g., empirical experiments in a scientific discovery context, or code execution in a debugging context) that returns a binary signal indicating whether the candidate $c$ is consistent with the target $c \subseteq c^{\star}$. The search concludes successfully if $c^{\star}$ is found and verified.

A candidate $c$ is a set of \textit{literals} $\ell_{i,v}$, where each literal expresses a single assertion about the target. Let $q_1,\dots,q_M$ denote the questions defining a task, with $V_1,\dots,V_M$ denoting their possible answer sets; when these have uniform size, we write $|V_i|=K$. Then we an define a literal as 
\[ \ell_{i,v} \equiv (q_i=v), \qquad \mathcal{L} = \{\ell_{i,v}: i\in\{1,\dots,M\},\ v\in V_i\}. \]
The first index identifies the question, while the second represents the proposed answer. 
% A candidate $c\subseteq\mathcal{L}$ is a partial or complete hypothesis containing at most one answer to each question. 
Each candidate branches into $b$ child candidates by adding an additional literal, e.g., a candidate $c_0 = \{\ell_{0,\alpha}\}$ can branch into $c_1 = \{\ell_{0,\alpha}, \ell_{1,\alpha}\}$ and $c_2 = \{\ell_{0,\alpha}, \ell_{1,\beta}\}$ assuming $\alpha,\beta \in V_1$ and $b=2$.

We assume a system of $N$ agents. Agents traverse the graph in $T$ rounds, indexed by $t = \{1, 2, ... ,T \}$, where each round consists of performing verification of at most $m$ leaf candidates (i.e., candidates located at level $t$) and pruning candidates that fail verification.
Agent $a_i$ can communicate their verification results with any neighboring agent $a_j$, and $a_j$ can use these results to prune candidates that $a_i$ has verified (see Section~\ref{sec:branching}).
We parameterize agent communication topology simply with a single communication degree parameter $d$, representing the number of other agents each can communicate with.
A schematic of a round is illustrated in Figure~\ref{fig:setup-round}.

\subsection{Modeling Search Success: Absorbing State Phase Transitions}
An \textit{absorbing-state phase transition} occurs when changing a system parameter determines whether activity persists or eventually stops permanently (enters the \textit{absorbing state}). 
They are typically characterized by a sharp, nonlinear transition from activity to absorbing state at a \emph{critical value} of the system parameter.

In our setting, we measure the search activity as a function of communication degree $\commdeg$, and the absorbing state corresponds to a ``solved search''.
We define this state by parameterizing search activity as the number of active incorrect search candidates $\Ainc$, summed over
all agents. The absorbing state is reached when $\Ainc$ decays toward zero: eventually the correct candidate will be all that's left, and verification of $c^{\star}$ is \emph{guaranteed} within the round budget $\rounds$. 
We wish to derive the critical communication degree
$\critdeg$: for $\commdeg < \critdeg$ the incorrect-candidate population grows over the search horizon, whereas for $\commdeg > \critdeg$ it contracts; sufficiently strong accumulated decay of wrong hypothesis drives the system toward the solved absorbing state.

\subsection{Search Task Taxonomy}
\label{sec:taxonomy}

Communication helps search when evidence produced by one agent can remove incorrect hypotheses that others are exploring. 
We argue that only certain types of search tasks have the property of \textit{evidence reusability}, leading us to introduce a taxonomy for search tasks (Figure~\ref{fig:search-structures}) inspired by parallel search, AND/OR search regimes, and branch-and-bound algorithms \citep{karp1993randomized,dechter2007and,lai1984anomalies}.
% Thus, $\critdeg$ is a function of the type of task, 
\input{taxonomy}

Needle and chain search (Figure~\ref{fig:needle-structure}, ~\ref{fig:chain-structure}) are characterized by a branching factor $\bagent = 1$: the population of incorrect candidates cannot grow.
Needle search is purely an OR search space in which candidates cannot be further decomposed, so there is no branching behavior, and refuting a candidate has no effect on other candidates.
For example, a wrong guess for an unknown password provides no information about untested guesses.
In chain search, each of $C$ experiments must be done in sequence, e.g., following an unknown sequence of pointers when each query reveals the next.
Needle search is coverage-limited, so scaling agents helps but communication won't; while chain search is depth-limited, so parallelism cannot help at all.

Hierarchical and factorizable search (Figures~\ref{fig:tree-structure}, ~\ref{fig:graph-structure}) introduce branching, but differ
in the scope of reusable evidence.
% Classical AND/OR search bounds distinguish tree exploration, whose complexity depends on decomposition depth, from search with shared subproblems, whose complexity only depends on subproblem size, which is often much smaller \citep{dechter2007and}.
In our coarse-to-fine hierarchical model, which represents tasks such as localizing a faulty component or buggy commit, candidates are specific to their parents, so an incorrect candidate can only prune its corresponding subtree.
In our factorizable model, on the other hand, literals recur across paths, so refuting one literal can eliminate candidates in other parts of the graph.
Unlike distributing a fixed computation across agents, communicating verification results between agents can reduce the total work required by subsequent search.
Parallel branch-and-bound provides a classical example: an improved solution discovered by one worker supplies a bound that allows others to prune unexplored subtrees, potentially producing superlinear speedup relative to a sequential search \citep{lai1984anomalies}.
Section~\ref{sec:branching} models this interaction between candidate proliferation and evidence-driven pruning.

\section{Theoretical Results}
\label{sec:theory}

We theoretically derive $\critdeg$ for all four search task classes. We show that needle and chain search tasks will never reach a phase transition in communication degree (Section~\ref{sec:no-pt}), whereas $\critdeg$ exists for tree and factorizable search (Section~\ref{sec:branching}).

\subsection{Search without phase transitions}
\label{sec:no-pt}

\subsubsection{Needle search}

Needle or unstructured search
(Figure~\ref{fig:needle-structure})
is characterized by unrelated candidates: excluding one candidate
provides no information about any others, and there is no persistent production of new
incorrect candidates.
Communication can reduce redundant verification, but there is no
branching--pruning balance and hence no nontrivial critical degree.
We denote this absence of a positive threshold by
\begin{equation}
d_c^{\mathrm{needle}}=0.
\end{equation}
\subsubsection{Chain search}

Chain or sequential search (Figure~\ref{fig:chain-structure})
requires $C$ experiments in sequence, where candidate $j+1$ becomes
available only after verification of candidate $j$ is complete.
Under the assumption that only one sequential experiment can be
completed per round, communication cannot shorten this dependency
chain. Thus,
\begin{equation}
d_c^{\mathrm{chain}}(C,T)=
\begin{cases}
0,      & T\geq C,\\
\infty, & T<C.
\end{cases}
\label{eq:chain}
\end{equation}
\subsection{Search with phase transitions}
\label{sec:branching}

Hierarchical and factorizable tasks share a balance between branching
and verification. Let $A_t$ denote the expected total number of active incorrect
candidates across all $N$ agents, counting copies separately.
With branching factor $b$, normalized budget parameter $m$
(corresponding to a total round budget $Nm$), and communication degree $d$,
the mean-field dynamics are
\begin{equation}
A_{t+1}\simeq r_t(d)(A_t-mN),
\qquad
r_t(d)=b(1-mq_t)^d,
\end{equation}
where $q_t$ is the probability that a single communicated
verification result eliminates a given incorrect candidate
at a receiving agent.
This approximation assumes that the full system-wide budget $Nm$
is used each round and allocated among eligible agents, and that
experiments are selected randomly and independently.

For $A_t\gg mN$, self-pruning is subleading, giving
$A_T\simeq A_1 e^{(T-1)\bar{\lambda}_T(d)}$, where
\begin{equation}
    \bar{\lambda}_T(d)\equiv
(T-1)^{-1}\sum_{t=1}^{T-1}\ln r_t(d).
\end{equation}
The finite-horizon critical degree, defined by
$\bar{\lambda}_T(d_c)=0$, is therefore
\begin{equation}
\label{eqn:dc-general}
d_c
=
-\frac{(T-1)\ln b}
{\sum\nolimits_{t=1}^{T-1}\ln(1-mq_t)}
\simeq
\frac{(T-1)\ln b}
{m\sum\nolimits_{t=1}^{T-1}q_t},
\end{equation}
where the approximation holds for $mq_t\ll1$ at every round.
Below $d_c$, incorrect candidates proliferate ($A_T>A_1$);
above $d_c$, they decay ($A_T<A_1$).
Under the additional idealization of uniform final selection from
these incorrect candidates and one retained correct candidate,
\begin{equation}
P_{\mathrm{succ}}(d)
\simeq \frac{1}{1+A_T(d)}
\simeq
\frac{1}{1+A_1e^{(T-1)\bar{\lambda}_T(d)}}.
\end{equation}
Success is below $1/(1+A_1)$ for $d<d_c$ and above it for $d>d_c$.
The limits $P_{\mathrm{succ}}\to0$ and $P_{\mathrm{succ}}\to1$
follow when the accumulated log reproduction
$(T-1)\bar{\lambda}_T(d)$ tends to $+\infty$ and $-\infty$,
respectively; finite horizons give a smooth crossover.

\subsubsection{Hierarchical search}

In hierarchical coarse-to-fine search, evidence transfers only
between candidates sharing the refuted path.
Under uniform, independent selection among $b^t$ depth-$t$ paths,
\begin{equation}
q_t=b^{-t},
\qquad
d_c\simeq
\frac{(T-1)(b-1)\ln b}
{m\left[1-b^{-(T-1)}\right]}.
\end{equation}
Here and below, we use the weak-pruning approximation
$\ln(1-mq_t)\simeq -mq_t$, valid when $mq_t\ll1$ at every round.
For hierarchical search, this requires $m/b\ll1$.
Cumulative overlap $\sum_{t=1}^{T-1}b^{-t}$ saturates at $1/(b-1)$, while branching continues
each round. As search deepens, agents spread across increasingly many branches, and their evidence becomes less mutually relevant.
Consequently, $d_c\propto T$ at large $T$:
deeper search requires more communication, but $d_c$ cannot exceed $N-1$.

% Because overlap decreases exponentially with depth, each additional round contributes less to evidence reuse, and the cumulative overlap $\sum_{t=1}^{T-1}b^{-t}$ approaches $1/(b-1)$ for $b>1$.
% Branching, however, continues at every round.
% Maintaining a balance therefore requires a communication degree that grows approximately linearly with $T$.
% For a fixed number of agents, the predicted threshold eventually exceeds the maximum available degree $N-1$: even communication with every other agent cannot offset branching under these assumptions.

\subsubsection{Factorizable search}

In factorizable search, a depth-$t$ candidate contains $t$ literals
from a pool $\mathcal{L}$.
Refuting a literal eliminates every candidate containing it,
regardless of its refinement path. Since a depth-$t$ candidate contains $t$ literals from a pool
$\mathcal L$, a uniformly sampled literal overlaps with it with probability
$t/|\mathcal L|$.
\begin{equation}
q_t\simeq\frac{t}{|\mathcal{L}|},
\qquad
\sum_{t=1}^{T-1}q_t
\simeq\frac{T(T-1)}{2|\mathcal{L}|},
\qquad
d_c\simeq\frac{2|\mathcal{L}|\ln b}{mT}.
\end{equation}
Unlike hierarchical search, cumulative overlap grows quadratically with $T$ while branching grows only linearly. 
Thus, $d_c\propto T^{-1}$: deeper candidates contain more components and thus offer more opportunities for pruning through reusable component-level evidence.

\section{Experiments}
\label{sec:exps}

We run empirical experiments on three factorizable search problems evaluate the applicability of our theoretically predicted critical communication degree $\critdeg$ to LLM agents.

\subsection{Tasks}

\paragraph{Simple toy example.}
Agents search for a hidden configuration of $T$ literals, each taking one of $\numvals$ values. A candidate is a partial configuration, and branching occurs by adding an additional literal (see Figure~\ref{fig:graph-structure}), meaning that a candidate at depth $t$ in the graph will have $t$ literals. A candidate passes verification if it is a subset of the target configuration. The task is solved when a candidate is verified to exactly match the hidden configuration.

\paragraph{Configuration fault localization (TCAS).} 
We test on a module of the Traffic Collision Avoidance System (TCAS) benchmark \citep{hutchins1994experiments,do2005supporting,kuhn2006pseudo}, which takes $M=12$ multi-valued inputs encoding aircraft altitude and separation data, to determine whether to issue an upward, downward, or no resolution advisory.
% The total input combination of the TCAS is 3 × 23 × 3 × 2 × 4 × 102 × 3 × 2 × 3 = 1036800, which we hypothesize can be substantially sped up through communicative multi-agent search.
Similar to the toy example, branching occurs through adding a literal, and the task is solved when a specific hidden configuration is discovered.

\paragraph{Physical mechanism discovery.}
We test on symbolic discovery of a dynamical law $\dot{z} = f^\star(z, t)$,
abstracting the interactive discovery setting of ODEBench
\citep{dascoli2024odeformer} and NewtonBench \citep{zheng2026newtonbench}. We constructed a simple dynamical-system discovery environment with a fixed library of 16 symbolic force terms $\numcands$ (e.g., $x$, $x^3$, $\sin x$, $\dot{x}$, $\cos(\omega t)$), from which four are sampled to define each hidden equation, and agents work off of the resulting physical trajectories.
Following SINDy-style support recovery \citep{brunton2016sindy}, the search target is the sparse subset $S^\star \subseteq \numcands$ of active terms.
Verification is testing for membership of a single term in $S^\star$, and the task is solved when $S^\star$ is discovered and verified.
% As in the toy example, a candidate at round $t$ consists of $t$ terms, so branching occurs by addingone further library term.

\subsection{Experimental Procedure}

\paragraph{Models.}
We evaluate deterministic non-LLM agents alongside agents based on
four LLMs spanning open-weight and closed-source models:
Qwen3.5-4B, Qwen3.5-9B, GPT-5.4 nano, and GPT-6 luna.
All models were evaluated with sampling temperature 1.
For GPT-6 luna, we additionally compare five reasoning-effort
settings: \texttt{none}, \texttt{low}, \texttt{medium},
\texttt{high}, and \texttt{xhigh}.

% \paragraph{Task complexity.} Explain $C_{\text{task}}$ here. \todo{do we still need this?}

\paragraph{Multi-agent search procedure.}
We use the same search protocol for synthetic, TCAS, and physics-discovery
tasks. 
Deterministic agent experiments are purely rule-based: a total round
budget of $Nm$ candidate verifications is allocated across agents with
eligible candidates each round. Inter-agent communication of verification results and pruning of incorrect branches occur automatically.
We relax some of these assumptions in our LLM experiments.
Agents decide which experiments to perform and how/what to prune
within the same total round budget $Nm$.
To probe the role of 1) cooperation-focused prompting, and 2) free vs. automatic communication of verification results, we consider three policies outlined in Table~\ref{tab:llm_policies}.

\paragraph{Experimental configurations.}
All benchmarks use $N=40$ agents, branching factor $b=2$, per-agent verification budget $m=1$, and each agent starts with $n_0=8$ candidates.
The task-specific search spaces have $|\mathcal{L}|$ valid literals and target depth $T$: $(|\mathcal{L}|,T)=(16,4)$ for the toy task, $(46,6)$ for TCAS, and $(16,4)$ for physical mechanism discovery.
For the toy task, the 16 literals arise from $M=8$ components with $K=2$.
For TCAS, the 12 parameters have heterogeneous value cardinalities, with $K_{\max}=10$ and 46 valid literals in total.
For mechanism discovery, we use $M=16$ candidate equation components with each literal defined as a binary inclusion choice.

\begin{table}[t]
\centering
\caption{LLM communication policies tested.}
\label{tab:llm_policies}
\begin{tabularx}{\linewidth}{@{}l l X@{}}
\toprule
Policy & Prompted objective & Communication rule \\
\midrule
Team-Auto & Team performance
   & Automatically transmit the true verdict of every experiment to $d$ neighbors (same as for deterministic agents). \\
Team-Free & Team performance
   & The agent chooses which of $d$ neighbors to communicate with, what to report (deception is allowed), plus optional text. \\
Indiv-Free & Individual performance
   & Same as C1. \\
\bottomrule
\end{tabularx}
\end{table}

\paragraph{Evaluation.}
Within each benchmark, all policies use the same set of communication degree $d \in \{0, 8, 16, 24, 39\}$ and six
task seeds. We estimate the cumulative incorrect-lineage reproduction
factor by pooling counts across agents and tasks:
\begin{equation}
\lambda(d)
=\sum_{t=1}^{T-1}
\log\left(\frac{\sum_e C_t^{(e)}(d)}{\sum_e A_{t-1}^{(e)}(d)}\right).
\end{equation}
Here $e$ indexes tasks, $A_{t-1}^{(e)}$ counts incorrect hypotheses at the
start of round $t$, and $C_t^{(e)}$ counts their children, deduplicated within
each agent, after pruning and branching.
Children descended only from correct parents are excluded from the numerator.
We estimate $d_c$ from the per-round average $\bar\lambda(d_c)=0$ by linearly interpolating $\bar\lambda (d)$ between consecutive sampled degrees.
% with finite values on opposite sides of zero.

\section{Results and Discussion}
\label{sec:results}

\subsection{Deterministic Models Agree With Theory}

\begin{figure}
    \centering
    \includegraphics[width=0.78\linewidth]{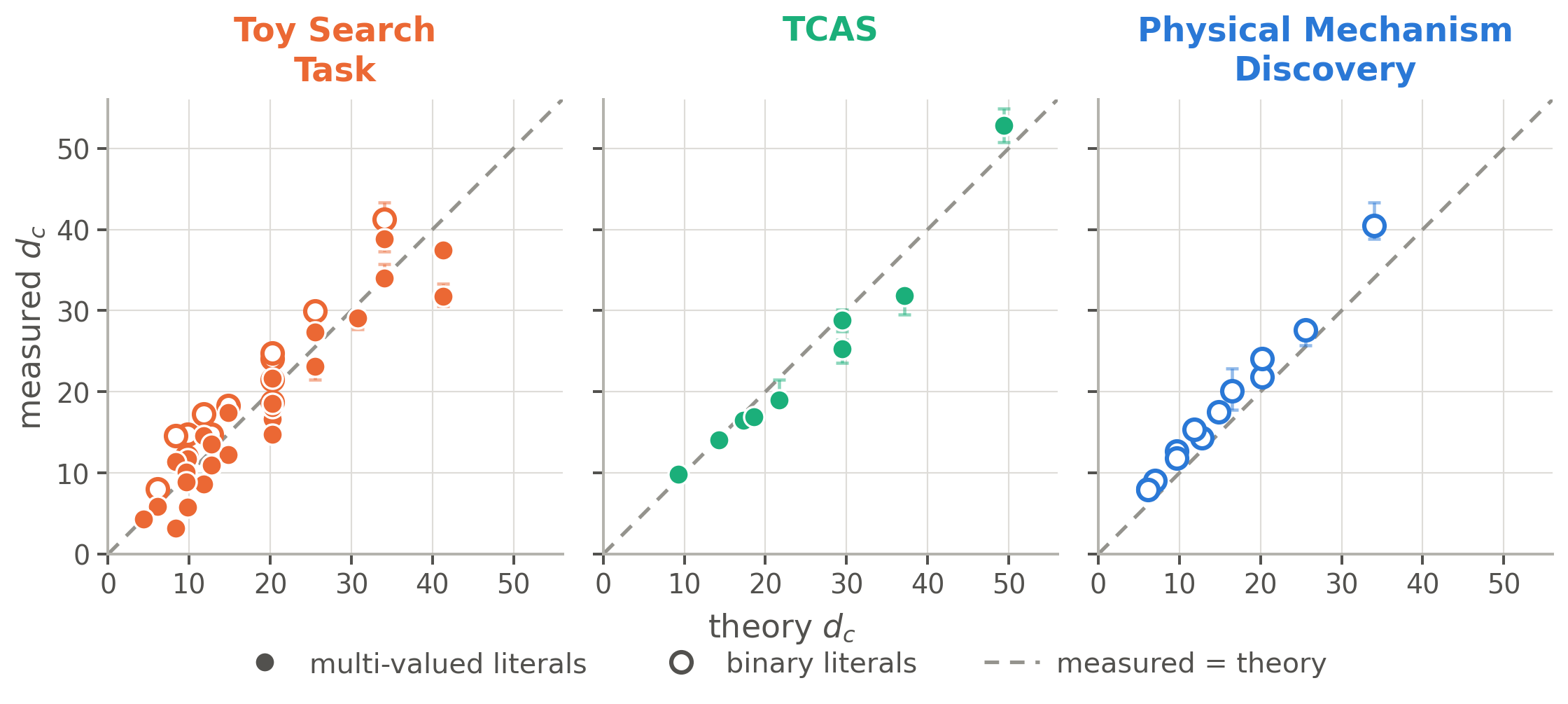}
    \caption{\textbf{Deterministic agents recover theoretical $d_c$ prediction across all tasks.} Filled points represent task configurations with multi-valued literals ($|V_i|>2$), such as in the TCAS task. Open points represent $|V_i|=2$, such as in the physical mechanism discovery task where each symbolic term either is or is not in the true equation.}
    \label{fig:deterministic}
\end{figure}

To ensure that our theory models real-world factorizable search problems, we evaluate deterministic multi-agent systems on our 3 factorizable search tasks.
We find very good agreement between theoretically predicted critical communication degree $d_c$ and measured $d_c$ from our experiments across all tasks (Figure~\ref{fig:deterministic}).
Measured $d_c$ tracks theory closely in all three environments (Lin's CCC = 0.93, 0.97 and 0.94 for the toy example, TCAS, and mechanism discovery; respectively). 

The mechanism discovery task shows the most disagreement, where measured $d_c$ exceeds theory by an average of 17\%.
We note that in this task, literals are binary since each candidate mechanism is either present or absent.
The literals in the TCAS task, on the other hand, take one of several mutually exclusive values; and we tested the toy task in both binary and multi-valued settings.
Settings with binary literals consistently produce higher $d_c$ than predicted by theory.

\subsection{Experiments on LLM Agents}

\paragraph{Comparison with theoretical prediction.}
% Our empirically measured $d_c$ values for all LLM model families and tasks can be found in Table~\ref{tbl:dc_results}.
Through these experiments, we wished to understand whether capable LLMs can discover strategies that are better than the theoretical setup, such as cooperation or intelligent candidate selection, and thus achieve a $d_c$ lower than theory.
While the best agent-policy combinations approach theoretical prediction, we do not observe a statistically clear improvement over theory in any tested case.
We find that agents tend to add literals that have already been validated, leaving previously introduced incorrect components untested, and overlapping experimental choices further produce redundant evidence. 

\begin{table}
    \centering
    \caption{\textbf{Empirical critical communication degrees $\widehat d_c$ for four LLM families.}
Estimates are obtained from pooled incorrect-lineage reproduction
counts over six task seeds; error bars denote 95\% task-level bootstrap
confidence intervals.
Lowest $\hat{d_c}$ values are bolded, and the theoretical prediction for $d_c$ for each task is shown below each task name for comparison.}

% ============ A: en-dash, 1 dp, scriptsize  (~5.36 in)
\begingroup\setlength{\tabcolsep}{3pt}
\begin{tabular*}{\textwidth}{@{\extracolsep{\fill}}l*{9}{c}@{}}
\toprule
& \multicolumn{3}{c}{Synthetic}
& \multicolumn{3}{c}{TCAS}
& \multicolumn{3}{c}{Physics} \\

& \multicolumn{3}{c}{$d_c^{\mathrm{theory}} = 6.10$}
& \multicolumn{3}{c}{$d_c^{\mathrm{theory}} = 10.90$}
& \multicolumn{3}{c}{$d_c^{\mathrm{theory}} = 6.10$} \\

\cmidrule(lr){2-4} \cmidrule(lr){5-7} \cmidrule(lr){8-10}

Model
& \makecell{Team-\\Auto}
& \makecell{Team-\\Free}
& \makecell{Indiv-\\Free}
& \makecell{Team-\\Auto}
& \makecell{Team-\\Free}
& \makecell{Indiv-\\Free}
& \makecell{Team-\\Auto}
& \makecell{Team-\\Free}
& \makecell{Indiv-\\Free} \\

\midrule

Qwen3.5-4B
& \makecell{6.55\\{\scriptsize 6.2--6.9}}
& \makecell{12.82\\{\scriptsize 12.2--13.4}}
& \makecell{13.47\\{\scriptsize 11.6--16.0}}
& \makecell{12.79\\{\scriptsize 11.4--13.7}}
& \makecell{17.69\\{\scriptsize 16.5--25.1}}
& \makecell{20.13\\{\scriptsize 16.8--24.8}}
& \makecell{8.77\\{\scriptsize 7.7--10.0}}
& \makecell{13.94\\{\scriptsize 12.2--14.6}}
& \makecell{14.31\\{\scriptsize 11.5--16.9}} \\

Qwen3.5-9B
& \makecell{6.64\\{\scriptsize 6.2--7.0}}
& \makecell{16.01\\{\scriptsize 13.8--17.4}}
& \makecell{16.06\\{\scriptsize 14.1--18.2}}
& \makecell{12.90\\{\scriptsize 10.9--15.2}}
& \makecell{29.15\\{\scriptsize 23.6--37.3}}
& \makecell{27.31\\{\scriptsize 24.8--29.6}}
& \makecell{8.34\\{\scriptsize 7.7--8.8}}
& \makecell{17.07\\{\scriptsize 15.0--18.9}}
& \makecell{16.77\\{\scriptsize 14.2--18.5}} \\

GPT-5.4 Nano
& \makecell{\textbf{6.39}\\{\scriptsize 5.8--7.1}}
& \makecell{7.87\\{\scriptsize 7.0--9.0}}
& \makecell{7.29\\{\scriptsize 6.8--7.8}}
& \makecell{10.52\\{\scriptsize 8.7--11.6}}
& \makecell{\textbf{12.88}\\{\scriptsize 9.8--14.5}}
& \makecell{14.41\\{\scriptsize 9.7--16.1}}
& \makecell{10.95\\{\scriptsize 9.8--12.0}}
& \makecell{13.04\\{\scriptsize 11.3--16.7}}
& \makecell{12.79\\{\scriptsize 11.8--13.4}} \\

GPT-6 Luna
& \makecell{6.88\\{\scriptsize 6.4--7.3}}
& \makecell{\textbf{7.06}\\{\scriptsize 6.8--7.3}}
& \makecell{\textbf{6.46}\\{\scriptsize 6.0--7.0}}
& \makecell{\textbf{10.05}\\{\scriptsize 9.6--11.2}}
& \makecell{13.71\\{\scriptsize 10.0--18.0}}
& \makecell{\textbf{13.13}\\{\scriptsize 9.3--14.9}}
& \makecell{\textbf{8.17}\\{\scriptsize 7.3--9.0}}
& \makecell{\textbf{8.48}\\{\scriptsize 7.9--9.2}}
& \makecell{\textbf{7.88}\\{\scriptsize 7.4--8.8}} \\

\bottomrule
\end{tabular*}
\endgroup

\label{tbl:dc_results}
\end{table}

\paragraph{Incomplete communication leads to higher $d_c$.}
While the GPT models have slightly lower $\hat{d_c}$ values compared to the Qwen models, the difference is most pronounced under free communication policies (Team-Free and Indiv-Free).
Free communication can further reduce effective pruning because agents often do not report all experimental outcomes, especially negative evidence that could eliminate incorrect hypotheses (Table~\ref{tab:comm-honesty}). We find that the relative increase in \(d_c\) under free communication, quantified by $d_c^{\mathrm{Team\text{-}Free}}/d_c^{\mathrm{Team\text{-}Auto}}$ is strongly negatively correlated ($r=-0.91$) with the rate at which agents report their experimental results (Figure~\ref{fig:reporting-rate}).

\begin{figure}
    \centering
    \includegraphics[width=0.88\linewidth]{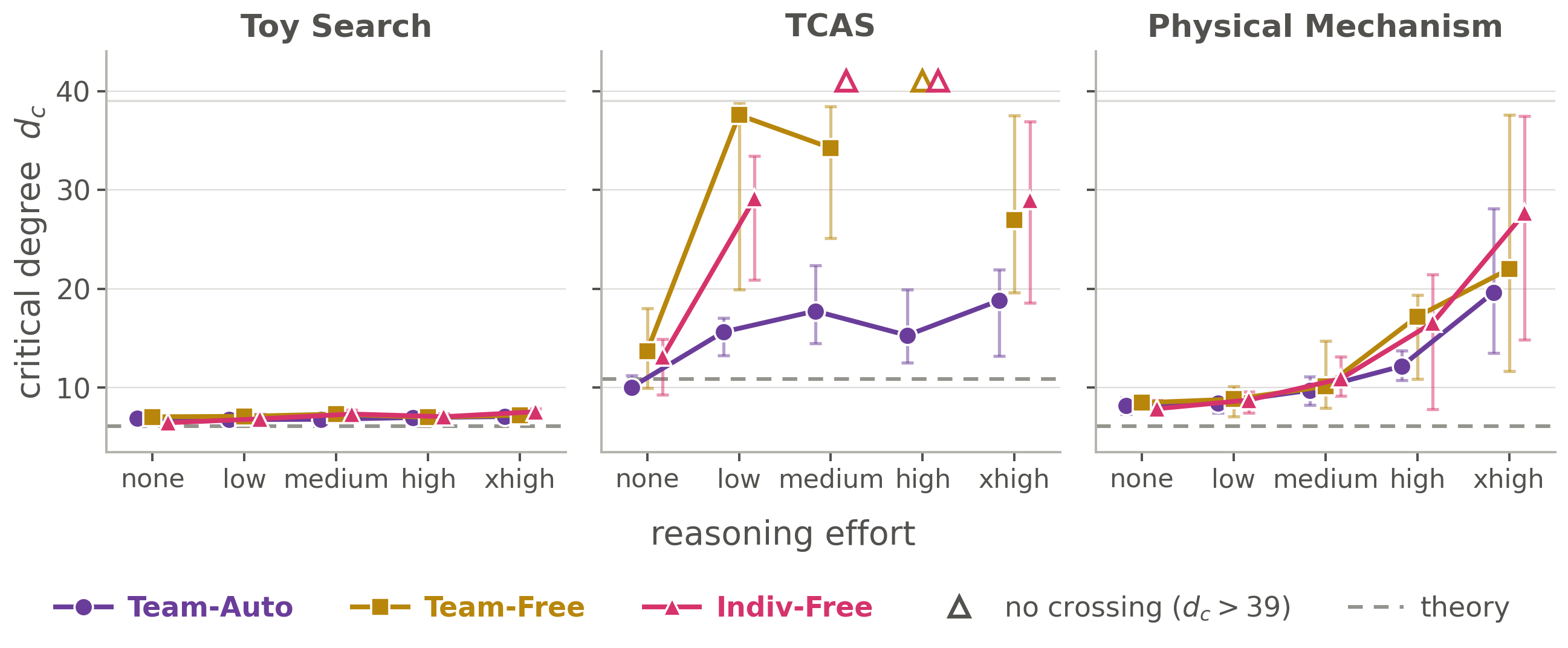}
    \caption{Higher reasoning effort on GPT-6-luna almost always raises the critical communication degree $d_c$ across all policies and tasks.}
    \label{fig:reasoning}
\end{figure}

\paragraph{Higher reasoning effort increases $d_c$.}
We ablate GPT-6-luna's reasoning effort, and find that higher reasoning effort either has minimal effect or actually increases the required $d_c$, moving the result further away from the theoretical prediction (Figure~\ref{fig:reasoning}).
We find that more reasoning capability led agents to herd into the most informative paths, but this decreases the diversity of explored paths and diminishes the realized benefits of communication. The average number of unique literals tested by agents decreases from $25.5 \rightarrow 11.3$ in the TCAS task, and $14.2 \rightarrow 8.8$ for the physics task.
However, agents \textit{individually} are either just as likely or more likely to complete the search task with higher reasoning (Figure~\ref{fig:phase-diag-reasoning}).

\begin{figure}
    \centering
    \includegraphics[width=0.9\linewidth]{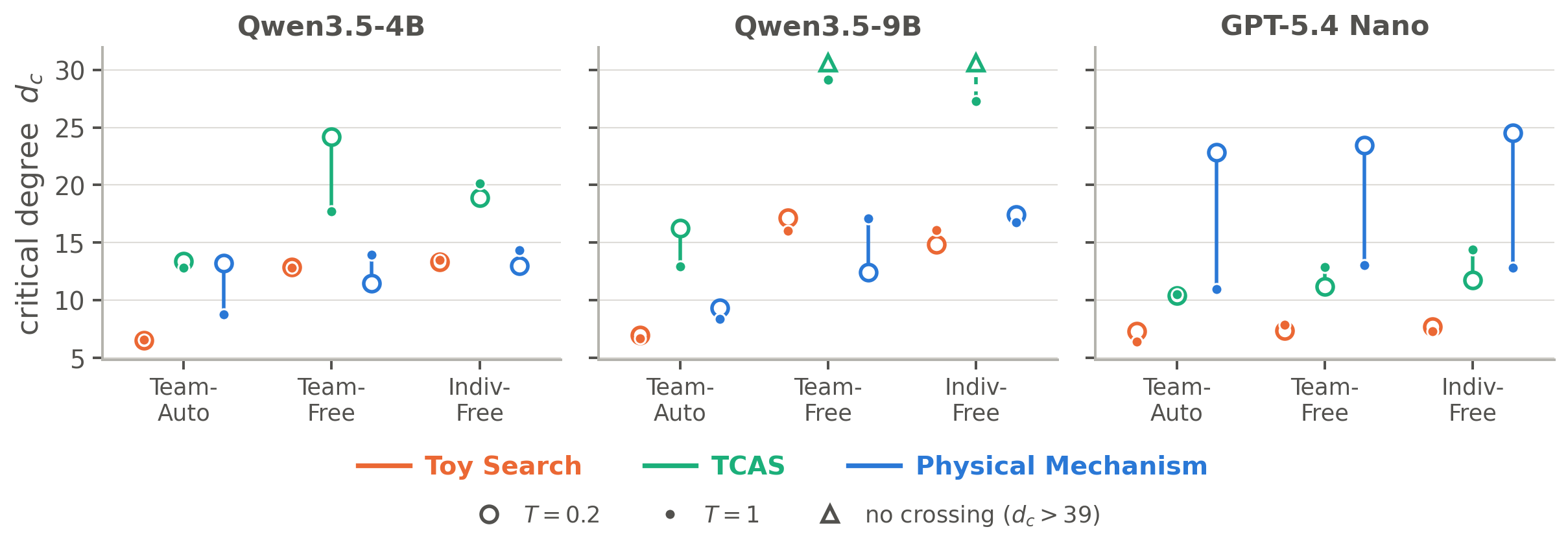}
    \caption{Changing the sampling temperature can produce very different $d_c$ for certain models and tasks. We test $T=0.2$ (filled point) against the default temperature setting $T=1$ (open circle) on all policies and tasks.}
    \label{fig:temperature}
\end{figure}

\textbf{Sampling temperature has model- and task-dependent effects.}
We test the effect of sampling temperature by comparing \(T=0.2\) with the default \(T=1\) for three of the four models, as \(T=0.2\) is not available for GPT-6 Luna (Figure~\ref{fig:temperature}).
We find that higher temperature makes agents select more wider range of candidates, allowing them to reap the benefits of communication.
$T=0.2$ produced a particularly noticeable $d_c$ increase for the physical mechanism discovery task for GPT-5.4-Nano specifically, while making relatively insignificant changes to the results of other tasks and model families.
This may be due to the inherent differences in task difficulty: e.g., adding a linear damping term is very obvious for the physical trajectory given, while few clues exist at the start when searching for a hidden configuration.

\section{Related Works}
\label{sec:rw}
    
\paragraph{Multi-agent search.}
A growing literature uses systems of LLM agents to improve search efficiency through coordination and parallelization.
Several works contribute orchestration strategies for multi-agent information seeking and evidence gathering \citep{song2026webswarm,prabhakar2025enterprisedeepresearch,chen2025mindsearch,jin2025hira} and benchmark such strategies \citep{wong2026widesearch}. The line of work most relevant to us is search over possible solutions \citep{yang2026mosa}, hidden transformations from input to output \citep{li2026matsir}, and autoresearch on ML algorithm performance optimization \citep{jin2026hypothesistree}.
Our work analyzes the dynamics of multi-agent search problems from a theoretical lens as well as contributes empirical observations from LLM-based systems.

\paragraph{Role of task structure and communication in multi-agent systems.}
The choice of communication topology in multi-agent systems has a significant impact on downstream task solving ability and efficiency \citep{kim2025scaling,zhu2025multiagentbench}, and substantial research effort has been dedicated to designing (or learning) the optimal topology for a given problem \citep[e.g.,][]{zhang2024gdesigner,zhuge2024gptswarm,jiang-etal-2026-dynamic,shen-etal-2025-understanding,zhou2026multi,zhou2026mass}.
Task structure and complexity also plays a major role in the success of the multi-agent system, and different tasks have been found to require different communication protocols \citep{rizvimartel2025benefits,kim2025scaling,tang2025taskcomplexity,qian2025macnet,yang2025topological}.
Our work contributes a theory-grounded taxonomy of search task types and investigates the dependence of communication degree on task type.

\paragraph{Collective dynamics and phase transitions in multi-agent systems.} 

Recent works have applied the machinery of statistical mechanics and phase transitions to model the dynamics of LLM multi-agent systems.
Existing works primarily leverage statistical physics theory to understand the formation of collective opinion or consensus in these systems \citep{el2026physicsagentsstatisticalmechanics,okawa2026biasedconsensus,denobili2026collectivealignment,ashery2025conventions,demarzo2026coordination,flint2026groupsize} or the propagation and amplification of errors through communication in multi-agent systems \citep{niu2026reliability,xie2026errorcascades}.
In contrast, our focuses on the problem of multi-agent search.

\paragraph{Classical foundations: cooperative search, absorbing state search theory.}
Our work directly builds off foundational work on absorbing state phase transitions \citep{hinrichsen2000absorbing} and its applications to search problems \citep{semerjian2003randomwalksat,lee2010finitesize}.
Our taxonomy of search tasks is informed by classical work on search task decomposition \citep{dechter2007and, dechter1986learning} and the role of evidence reusability \citep{clearwater1991cooperative,hamadi2009manysat}, as well as parallel branch-and-bound search \citep{lai1984anomalies,marinescu2009andor}.

\section{Conclusion}
We developed an absorbing-state framework for multi-agent search that connects task structure and evidence reuse to the critical communication degree required to suppress the proliferation of incorrect candidates. Experiments on three factorizable search tasks show close agreement with theory for deterministic rule-based agents, while LLM agents exhibit deviations associated with incomplete reporting and redundant exploration. These findings motivate designing communication protocols jointly with agents' search strategies, so that shared evidence supports both effective pruning and diverse exploration.

\subsection*{Acknowledgmments}
This work was supported by the National Science Foundation (NSF) under grant number 2341055. The Flatiron Institute is funded by the Simons Foundation.

\bibliography{addl_refs,references}
\bibliographystyle{plainnat}
\newpage
\appendix
\section{Additional Theoretical Analysis}

\subsection{Concentration of Agents}
\label{app:concentration of agents}

Using the same notation as in Section~\ref{sec:branching}, the
round-dependent mean-field dynamics are
\[
A_{t+1}
\simeq
r_t^{\rm MF}(\commdeg)
\left(A_t-\numexp\numagents\right),
\]
with
\[
r_t^{\rm MF}(\commdeg)
=
\bagent
\left(1-\numexp q_t\right)^{\commdeg}.
\]
Here the pruning and overlap probabilities are evaluated at step $t$.
As in the main text, $\Ainc[t]$ counts active incorrect candidates
across all agents, with copies held by different agents counted
separately.

This description assumes sufficient experiment supply and
approximately homogeneous, independent pruning opportunities from
different neighbors.
Initialization can affect both assumptions through the width and
composition of the local hypothesis frontiers.
We distinguish two possible sources of initialization sensitivity:
frontier extinction, which reduces experiment supply, and
correlated or redundant evidence, which reduces the pruning value
of communication.
These mechanisms provide possible explanations for deviations from
mean-field predictions; a shift in the critical degree alone does
not identify either mechanism.

\subsubsection{Frontier extinction}

An agent with a narrow hypothesis frontier may lose all of its
surviving branches after receiving valid evidence, even when a large
unexplored hypothesis space remains globally.
Without frontier replenishment, the agent can no longer propose
experiments from surviving hypotheses.
Communication can therefore reduce the subsequent supply of
evidence through the sequence
\[
\begin{aligned}
\text{incoming evidence}
&\;\Rightarrow\;
\text{local pruning}
\;\Rightarrow\;
\text{frontier extinction}
\\
&\;\Rightarrow\;
\text{candidate depletion}
\;\Rightarrow\;
\text{budget underuse}.
\end{aligned}
\]
Initial frontier width and composition affect the number of
alternative branches available after early pruning events.
This mechanism does not require different agents to occupy the
same hypothesis: distinct local frontiers can be exhausted
independently.
Moreover, an empty frontier does not itself certify task success,
since the correct branch may also be absent.

Define the budget-utilization fraction and active-agent fraction by
\[
u_t
=
\frac{B_t^{\rm used}}{\numagents\numexp},
\qquad
a_t
=
\frac{N_t^{\rm act}}{\numagents},
\]
where $B_t^{\rm used}$ is the number of experiments performed in
round $t$, and $N_t^{\rm act}$ counts agents with at least one
eligible experiment immediately before experiment allocation. When the total budget can be redistributed among active agents,
$a_t<1$ does not necessarily imply $u_t<1$.
The remaining agents may absorb the unused allocations, provided
their candidate capacity is sufficient.

To account for reduced experiment supply, we define
\[
q_{t,\rm eff}
\equiv
u_{t-1}q_t,
\]
so that the effective communication-induced reproduction factor becomes
\[
r_{t,\rm eff}(\commdeg)
=
\bagent
\left(1-\numexp u_{t-1}q_t\right)^{\commdeg}.
\]
The index $t-1$ reflects one-round delayed communication; an initially
empty inbox contributes no communication pruning. The mean-field supply
assumption corresponds to $u_{t-1}\simeq 1$. When $u_{t-1}<1$, communication
provides less pruning than predicted by mean field, shifting the
critical degree toward
\[
d_c^{\rm empirical}>d_c^{\rm MF}.
\]

\subsubsection{Correlated evidence}

Even when experiment supply is sufficient, the assumption of
independent neighbor evidence can fail at large communication degree.
Different agents may receive strongly overlapping sets of evidence.
Their exact hypothesis frontiers need not become similar, but their
accumulated constraints can become correlated. Consequently,
different neighbors may send duplicate or already implied evidence.

The possible feedback is summarized schematically as
\begin{equation}
\begin{aligned}
\commdeg\uparrow
&\;\Rightarrow\;
\text{shared evidence}\uparrow
\;\Rightarrow\;
\text{constraint overlap}\uparrow
\\
&\;\Rightarrow\;
\text{redundant evidence}\uparrow
\;\Rightarrow\;
\text{communication efficiency}\downarrow.
\end{aligned}
\end{equation}
Duplicate observations alone do not establish this mechanism, since
independent experiments can also coincide.

Let $s_{\rm comm}(t,\commdeg)$ be the survival probability of an
incorrect candidate present immediately before communication. We define
an effective communication degree through
\[
s_{\rm comm}(t,\commdeg)
=
\left(1-\numexp q_t\right)^{
d_{\rm eff}(t,\commdeg)}.
\]
When the reference probability is accurate and neighbor pruning is
approximately homogeneous and independent,
$d_{\rm eff}\simeq\commdeg$.
Redundancy can instead give $d_{\rm eff}<\commdeg$.
This inequality is not universal: $d_{\rm eff}$ is a descriptive
diagnostic and can also reflect inaccuracies in the reference
probability or heterogeneity among candidates.

Holding the remaining reproduction dynamics fixed, weaker pruning
can shift the candidate-growth crossing toward
\[
d_c^{\rm empirical}>d_c^{\rm MF}.
\]

\subsection{Parent-locked Reference}
\label{app:parent_locked_reference}
The parent-locked regime remains useful as an analytically tractable
reference, but it should not be interpreted as the typical state of the
synthetic system. In this limit all active agents share the same parent
hypothesis at the beginning of a round,
\[
H_t^{(1)}
=
H_t^{(2)}
=
\cdots
=
H_t^{(N)},
\]
while independently sampling children from that common parent.

At depth $t$, the number of possible children is
\[
S_t=K(M-t+1).
\]
If an active agent generates $b$ children, the probability that a
specific child $h$ is generated is
\[
P(h\in\mathcal B_t^{(j)})
=
\frac{b}{K(M-t+1)}.
\]

Assume that the total experimental budget $Nm$ can be fully used and
redistributed among active agents. In the regime where its candidate capacity is
sufficient to use this allocation,
\[
P(\text{$h$ tested}\mid
h\in\mathcal B_t^{(j)},j\ {\rm active})
\simeq
\frac{m}{a_tb}.
\]

A randomly selected neighbor is active with probability $a_t$.
Therefore, the pruning probability contributed by one nominal neighbor is
\[
m q_t^{\rm lock}
=
a_t
\frac{b}{K(M-t+1)}
\frac{m}{a_t b}
=
\frac{m}{K(M-t+1)},
\]
The active fraction and branching number cancel as long as the total
experimental budget can indeed be used.

For $T\ll M$ and the standard full-depth approximation, this gives the
scaling
\[
d_c^{\rm lock}
\simeq
\frac{MK\ln b}{m}=\frac{|\mathcal{L}|\ln b}{m}.
\]

The parent-locked limit should therefore be regarded as a reference
regime in which agents share the same parent but retain independent
child sampling. 

In the concentrated regime, the critical communication degree is bounded by
\[
\frac{2|\mathcal{L}|\ln b}{mT}
\lesssim
d_c^{\rm conc}
\lesssim
\frac{|\mathcal{L}|\ln b}{m}.
\]

\section{Task Details}
\label{app:task_details}
We consider three tasks: synthetic component search, TCAS configuration-fault
localization, and physical mechanism discovery. All use the same
branching-and-pruning protocol, but differ in the meaning of a hypothesis,
the experimental interface, and the evidence available for pruning.

% \paragraph{Shared search and communication protocol.}
% There are $N$ agents connected by a fixed, undirected random $d$-regular
% graph, with $0\leq d\leq N-1$. Each agent maintains a private
% \emph{hypothesis frontier}: its set of surviving candidate hypotheses.
% Each round follows
% \[
% \text{receive/prune}
% \;\rightarrow\;
% \text{branch}
% \;\rightarrow\;
% \text{experiment/prune}
% \;\rightarrow\;
% \text{communication}.
% \]
% Each surviving hypothesis generates up to $b$ legal refinements, each
% adding one component without revising previous assignments. The total
% experimental budget is $B_{\mathrm{tot}}=Nm$ per round and is redistributed
% among agents that remain active, so $m$ is a nominal per-agent budget rather
% than a strict individual limit. Local evidence is applied immediately,
% whereas experimental records produced in round $t$ arrive at neighbors in
% round $t+1$. Agents communicate experimental records rather than their
% complete hypothesis frontiers or trees.

Throughout, $H^\star$ denotes the hidden mechanism, $T$ its number of
components, and $h_t$ a hypothesis containing $t$ components. Hypotheses
are sets, so different refinement orders can reach the same node. The
examples below are therefore partial views of a \emph{confluent hypothesis
graph}; blank nodes and ellipses indicate omitted hypotheses.

% ============================================================
\subsection{Synthetic Search Task}
\label{app:toy_task}

\paragraph{Hidden mechanism and hypotheses.}
The task has $M$ candidate components, each with a value set $\mathcal V$
of size $K$. We write $\ell_{j,a}$ for the atomic assignment
``component $j$ takes value $a$,'' where $j\in\{1,\ldots,M\}$ and
$a\in\mathcal V$. The hidden mechanism is
\[
H^\star=\{\ell_{j_1,a_1^\star},\ldots,\ell_{j_T,a_T^\star}\},
\qquad T\leq M,
\]
with distinct component indices $j_1,\ldots,j_T$. These indices are sampled
uniformly from the size-$T$ subsets of $\{1,\ldots,M\}$, and their values
are sampled uniformly from $\mathcal V$. A depth-$t$ hypothesis contains
$t$ assignments to distinct components; there are
$\binom{M}{t}K^t$ possible hypotheses at depth $t$. A partial hypothesis
is truth-compatible exactly when $h_t\subseteq H^\star$.

\paragraph{Experiments and evidence.}
An experiment tests one proposed atomic assignment $\ell_{j,a}$.
A refuted assignment prunes every local hypothesis containing it and is
not regenerated; communicated experimental records allow neighbors to
apply the same pruning rule.
A run succeeds when at least one agent uniquely identifies $H^\star$
at depth $T$.

\paragraph{Success criterion.}
A run succeeds when at least one agent identifies the complete
ground-truth mechanism $H^\star$ at depth $T$ and its constituent
assignments are supported by the accumulated experimental evidence.

\begin{figure}[H]
\centering
\resizebox{\linewidth}{!}{%
\begin{tikzpicture}[
    >=Stealth,
    mynode/.style={
        draw,
        rounded corners,
        align=center,
        font=\small,
        inner sep=4pt,
        minimum height=0.65cm,
        minimum width=1.1cm
    },
    note/.style={font=\scriptsize, align=center}
]

% depth labels
\node[font=\small] at (-9.2, 0)   {depth 0};
\node[font=\small] at (-9.2,-1.8) {depth 1};
\node[font=\small] at (-9.2,-3.8) {depth 2};
\node[font=\small] at (-9.2,-6.0) {depth 3};

% depth 0
\node[mynode] (root) at (0,0) {$\varnothing$};

% depth 1
\node[mynode] (a1) at (-4.5,-1.8) {$\ell_{1,1}$};
\node[mynode] (a2) at (0,-1.8)    {$\ell_{2,0}$};
\node[mynode] (a3) at (4.5,-1.8)  {$\ell_{3,1}$};
\node at (6.0,-1.8) {$\cdots$};

% depth 2
\node[mynode] (b1)  at (-6.6,-3.8) {$\ell_{1,1}\ell_{3,0}$};
\node[mynode] (b2)  at (-4.4,-3.8) {$\ell_{1,1}\ell_{3,1}$};
\node[mynode] (b12) at (-1.8,-3.8) {$\ell_{1,1}\ell_{2,0}$};
\node[mynode] (b3)  at (1.8,-3.8)  {$\ell_{2,0}\ell_{3,1}$};
\node[mynode] (b4)  at (3.4,-3.8)  {};
\node at (4.3,-3.8) {$\cdots$};
\node[mynode] (b5)  at (5.2,-3.8)  {};
\node[mynode] (b6)  at (6.6,-3.8)  {};
\node at (7.6,-3.8) {$\cdots$};

% depth 3
\node[mynode] (c1) at (-4.0,-6.0)
{$\ell_{1,1}\ell_{2,0}\ell_{4,0}$};

\node[mynode] (c2) at (-1.8,-6.0)
{$\ell_{1,1}\ell_{2,0}\ell_{3,1}$};

\node[mynode] (c3) at (0.4,-6.0)
{$\ell_{1,1}\ell_{2,0}\ell_{4,1}$};

\node at (1.7,-6.0) {$\cdots$};

% arrows: depth 0 -> 1
\draw[->] (root) -- (a1);
\draw[->] (root) -- (a2);
\draw[->] (root) -- (a3);

% arrows: depth 1 -> 2
\draw[->] (a1) -- (b1);
\draw[->] (a1) -- (b2);
\draw[->] (a1) -- (b12);
\draw[->] (a2) -- (b12);
\draw[->] (a2) -- (b3);
\draw[->] (a2) -- (b4);
\draw[->] (a3) -- (b5);
\draw[->] (a3) -- (b6);

% arrows: depth 2 -> 3
\draw[->] (b12) -- (c1);
\draw[->] (b12) -- (c2);
\draw[->] (b12) -- (c3);

% annotations
\node[note, text width=1.8cm] at (0,-3.9)
  {same hypothesis up to order $=$ same node};
\node[note] at (-6.6,-4.75) {experiment:\\ test if correct};
\node[note] at (-1.8,-6.95) {experiment:\\ test if correct};

\end{tikzpicture}}
\caption{
\textbf{Synthetic hypothesis search graph.}
Each symbol $\ell_{j,a}$ denotes the atomic assignment that component $j$
takes value $a$. Each refinement adds one previously unassigned
component--value pair. Different assignment orders can reach the same
hypothesis node, and experimentally refuted assignments prune all
hypotheses that contain them.
}
\label{fig:toy_hypothesis_tree}
\end{figure}

% ============================================================
\subsection{TCAS Configuration-Fault Localization}
\label{app:tcas_task}

\paragraph{Hidden mechanism and hypotheses.}
We use the configuration space of \texttt{tcas} from the Siemens test suite
as a controlled fault-localization benchmark
\citep{hutchins1994experiments,ghandehari2013applying}.
There are $M=12$ categorical parameters. Parameter $p_j$ has a legal value
set $\mathcal V_j$ of size $K_j$. The parameter cardinalities are seven
binary, two ternary, one four-valued, and two ten-valued, giving
\[
|\mathcal X|=\prod_{j=1}^{12}K_j=460{,}800,
\qquad
S=\sum_{j=1}^{12}K_j=44.
\]
Here $|\mathcal X|$ counts complete configurations, whereas $S$ counts
atomic parameter--value assignments. The hidden mechanism $H^\star$ is a
failure-inducing combination of $T$ assignments to distinct parameters.
A depth-$t$ hypothesis is
\[
h_t=\{p_{j_1}=v_{a_1},\ldots,p_{j_t}=v_{a_t}\},
\]
and each refinement assigns one previously unassigned parameter.
Different assignment orders can therefore reach the same partial
configuration, as illustrated in Fig.~\ref{fig:tcas_hypothesis_tree}.

\paragraph{Experiments and evidence.}
An experiment tests a complete configuration
$x=(x_1,\ldots,x_M)\in\mathcal X$ compatible with a proposed refinement
and returns $y(x)\in\{\mathrm{pass},\mathrm{fail}\}$.
We write $h\subseteq x$ when $x$ satisfies every assignment in $h$.
The configuration--outcome table can be precomputed, so an experimental
query is implemented as a table lookup. Under the conjunctive fault model,
a passing configuration rules out every candidate sufficient
failure-inducing combination contained in it:
\[
y(x)=\mathrm{pass},\quad h\subseteq x
\quad\Longrightarrow\quad
h\text{ is ruled out as a sufficient failure condition}.
\]
A failing configuration does not identify which subset of assignments
caused the failure and therefore does not, by itself, refute a particular
atomic assignment. Agents communicate the records $(x,y(x))$, allowing
neighbors to apply the same evidence to their own candidate hypotheses.
Unlike the synthetic component test, this experiment therefore returns a
verdict on a complete configuration rather than on one atomic assignment.

\paragraph{Success criterion.}
A complete candidate $h$ predicts failure when $h\subseteq x$ and pass
otherwise; denote this prediction by $\widehat y_h(x)$. It is evaluated
on a separate held-out configuration set $\mathcal E_{\mathrm{test}}$ using
\[
D(h,H^\star)=\frac{1}{|\mathcal E_{\mathrm{test}}|}
\sum_{x\in\mathcal E_{\mathrm{test}}}
\mathbf 1\!\left[\widehat y_h(x)\neq y^\star(x)\right],
\]
where $y^\star(x)$ is the hidden system's outcome.
A run succeeds when at least one agent reaches a depth-$T$ hypothesis
with $D(h,H^\star)<\epsilon$ that is also consistent with the
configuration--outcome evidence accumulated during search.

\begin{figure}[H]
\centering
\resizebox{\linewidth}{!}{%
\begin{tikzpicture}[
    >=Stealth,
    mynode/.style={
        draw,
        rounded corners,
        align=center,
        font=\small,
        inner sep=4pt,
        minimum height=0.65cm,
        minimum width=1.1cm
    },
    note/.style={font=\scriptsize, align=center}
]

% depth labels
\node[font=\small] at (-10.0, 0)   {depth 0};
\node[font=\small] at (-10.0,-1.8) {depth 1};
\node[font=\small] at (-10.0,-3.8) {depth 2};
\node[font=\small] at (-10.0,-6.0) {depth 3};

% depth 0
\node[mynode] (root) at (0,0) {$\varnothing$};

% depth 1
\node[mynode] (a1) at (-5.0,-1.8) {$p_1=v_1$};
\node[mynode] (a2) at (0,-1.8)    {$p_3=v_2$};
\node[mynode] (a3) at (5.0,-1.8)  {$p_7=v_1$};
\node at (6.8,-1.8) {$\cdots$};

% depth 2
\node[mynode] (b1)  at (-7.2,-3.8)
{$p_1=v_1,\;p_6=v_3$};

\node[mynode] (b2)  at (-4.2,-3.8)
{$p_1=v_1,\;p_9=v_4$};

\node[mynode] (b12) at (-0.2,-3.8)
{$p_1=v_1,\;p_3=v_2$};

\node[mynode] (b3)  at (2.9,-3.8)
{$p_3=v_2,\;p_7=v_1$};

\node[mynode] (b4) at (5.8,-3.8) {};
\node at (6.9,-3.8) {$\cdots$};

\node[mynode] (b5) at (8.3,-3.8) {};
\node[mynode] (b6) at (9.8,-3.8) {};
\node at (11.0,-3.8) {$\cdots$};

% depth 3
\node[mynode] (c1) at (-3.8,-6.0)
{$p_1=v_1,\;p_3=v_2,\;p_6=v_3$};

\node[mynode] (c2) at (0.4,-6.0)
{$p_1=v_1,\;p_3=v_2,\;p_8=v_1$};

\node[mynode] (c3) at (4.8,-6.0)
{$p_1=v_1,\;p_3=v_2,\;p_{10}=v_2$};

% arrows: depth 0 -> 1
\draw[->] (root) -- (a1);
\draw[->] (root) -- (a2);
\draw[->] (root) -- (a3);

% arrows: depth 1 -> 2
\draw[->] (a1) -- (b1);
\draw[->] (a1) -- (b2);
\draw[->] (a1) -- (b12);
\draw[->] (a2) -- (b12);
\draw[->] (a2) -- (b3);
\draw[->] (a2) -- (b4);
\draw[->] (a3) -- (b5);
\draw[->] (a3) -- (b6);

% arrows: depth 2 -> 3
\draw[->] (b12) -- (c1);
\draw[->] (b12) -- (c2);
\draw[->] (b12) -- (c3);

\end{tikzpicture}}
\caption{
\textbf{TCAS configuration-fault hypothesis graph.}
Each child adds one parameter--value assignment to its parent.
Different assignment orders that produce the same partial configuration
correspond to the same hypothesis node. Blank nodes and ellipses indicate
additional hypotheses omitted for clarity.
}
\label{fig:tcas_hypothesis_tree}
\end{figure}

% ============================================================
\subsection{Physical Mechanism Discovery Task}
\label{app:physics_task}

\paragraph{Hidden mechanism and hypotheses.}
Following the dynamical-system discovery setting of ODEBench and the
interactive probing setting of NewtonBench
\citep{dascoli2024odeformer,zheng2026newtonbench},
we consider a hidden equation of motion
\[
\ddot{x}=f^\star(x,\dot{x},\tau)
=\sum_{c\in H^\star}c(x,\dot{x},\tau),
\qquad |H^\star|=T.
\]
Here $\tau$ denotes physical time, to distinguish it from search depth $t$.
Candidate terms come from $M$ symbolic component families, each with $K$
allowed parameter choices. An atomic component specifies both a symbolic
family and its parameter choice. Examples include a linear restoring force,
a cubic restoring force, linear or state-dependent damping, and periodic
forcing. A depth-$t$ hypothesis $h_t=\{c_1,\ldots,c_t\}$ contains $t$
components, and a refinement adds one new symbolic component. Addition
order does not affect the resulting mechanism, as illustrated in
Fig.~\ref{fig:physics_hypothesis_tree}.

\paragraph{Experiments and evidence.}
Before the first communication round, each agent receives a different short
trajectory from the same hidden system. These local observations probe
different regions of state space and are held fixed across communication
degrees. An experiment selects an initial state
$z_0=(x_0,\dot{x}_0)$ and starting time $\tau_0$ from a fixed experiment
bank and returns
\[
\mathcal E(z_0,\tau_0)
=
\{(\tau_\ell,z(\tau_\ell))\}_{\ell=1}^{L},
\qquad
\tau_\ell\in[\tau_0,\tau_0+\Delta\tau],
\]
where $z=(x,\dot{x})$. The observation duration $\Delta\tau$, number of
measurements $L$, and measurement precision are fixed. Experiments are
chosen to help distinguish a proposed component from the current hypothesis.

In the controlled rule-based version, experimental evidence is reduced to
the idealized component-level outcome
\[
y(c)=\mathbf 1[c\in H^\star].
\]
This is the stipulated evidence interface for the controlled experiment,
not a claim that an arbitrary short trajectory uniquely identifies a term.
Negative evidence removes every hypothesis containing $c$; positive
evidence confirms $c$ and excludes incompatible parameter choices in the
same family. The testing agent keeps the physical trajectory private and
communicates only the component-level record $(c,y(c))$.

\paragraph{Success criterion.}
A run succeeds when at least one agent discovers a depth-$T$ mechanism
that is consistent with the experimental evidence accumulated during
search and predicts a separate set of held-out trajectories with error
\[
D(h,H^\star)<\epsilon.
\]
Here $D$ denotes held-out trajectory prediction error, distinct from the
configuration-classification error used for TCAS.

\begin{figure}[H]
\centering
\resizebox{\linewidth}{!}{%
\begin{tikzpicture}[
    >=Stealth,
    mynode/.style={
        draw,
        rounded corners,
        align=center,
        font=\small,
        inner sep=4pt,
        minimum height=0.65cm,
        minimum width=1.1cm
    },
    note/.style={font=\scriptsize, align=center}
]

% depth labels
\node[font=\small] at (-10.0, 0)   {depth 0};
\node[font=\small] at (-10.0,-1.8) {depth 1};
\node[font=\small] at (-10.0,-3.8) {depth 2};
\node[font=\small] at (-10.0,-6.0) {depth 3};

% depth 0
\node[mynode] (root) at (0,0) {$\varnothing$};

% depth 1
\node[mynode] (a1) at (-5.0,-1.8) {$-ax$};
\node[mynode] (a2) at (0,-1.8)    {$-bx^3$};
\node[mynode] (a3) at (5.0,-1.8)  {$-c\dot{x}$};
\node at (6.8,-1.8) {$\cdots$};

% depth 2
\node[mynode] (b1) at (-7.0,-3.8)
{$-ax,\,-c\dot{x}$};

\node[mynode] (b2) at (-4.0,-3.8)
{$-ax,\,+A\cos(\omega t)$};

\node[mynode] (b12) at (-0.3,-3.8)
{$-ax,\,-bx^3$};

\node[mynode] (b3) at (2.7,-3.8)
{$-bx^3,\,-c\dot{x}$};

\node[mynode] (b4) at (5.5,-3.8) {};
\node at (6.6,-3.8) {$\cdots$};

\node[mynode] (b5) at (8.1,-3.8) {};
\node[mynode] (b6) at (9.6,-3.8) {};
\node at (10.8,-3.8) {$\cdots$};

% depth 3
\node[mynode] (c1) at (-4.2,-6.0)
{$-ax,\,-bx^3,\,-c\dot{x}$};

\node[mynode] (c2) at (0.8,-6.0)
{$-ax,\,-bx^3,\,-\dfrac{c\dot{x}}{1+dx^2}$};

\node[mynode] (c3) at (5.4,-6.0)
{$-ax,\,-bx^3,\,+A\cos(\omega t)$};

% arrows: depth 0 -> 1
\draw[->] (root) -- (a1);
\draw[->] (root) -- (a2);
\draw[->] (root) -- (a3);

% arrows: depth 1 -> 2
\draw[->] (a1) -- (b1);
\draw[->] (a1) -- (b2);
\draw[->] (a1) -- (b12);
\draw[->] (a2) -- (b12);
\draw[->] (a2) -- (b3);
\draw[->] (a2) -- (b4);
\draw[->] (a3) -- (b5);
\draw[->] (a3) -- (b6);

% arrows: depth 2 -> 3
\draw[->] (b12) -- (c1);
\draw[->] (b12) -- (c2);
\draw[->] (b12) -- (c3);

\end{tikzpicture}}
\caption{
\textbf{Physical-mechanism hypothesis graph.}
Each child adds one atomic symbolic component to its parent.
Different addition orders can reach the same mechanism. Blank nodes and
ellipses indicate additional hypotheses omitted for clarity.
}
\label{fig:physics_hypothesis_tree}
\end{figure}

\section{Additional Results}

\subsection{Sharpening of the Success Crossover}
\label{app:success_sharpening}

We test whether the success crossover becomes sharper as the search horizon
$T$ increases. We use the synthetic benchmark and vary only $T$, while fixing
\[
N=80,\qquad M=8,\qquad K=4,\qquad b=3,\qquad m=1.
\]

For a given communication degree $d$, we define an episode as successful if
\emph{at least one agent recovers the true hypothesis with supporting evidence}.
The empirical success probability is therefore
\begin{equation}
P_{\mathrm{succ}}(d)
=
\frac{1}{E}
\sum_{e=1}^{E}
\mathbf{1}
\!\left[
\text{at least one agent succeeds in episode }e
\right],
\label{eq:empirical_success_probability}
\end{equation}
where $E=8$ episodes are evaluated at each degree. Consequently, the empirical
probabilities take values in increments of $1/8$.

\begin{figure}[H]
    \centering
    \includegraphics[width=\linewidth]
    {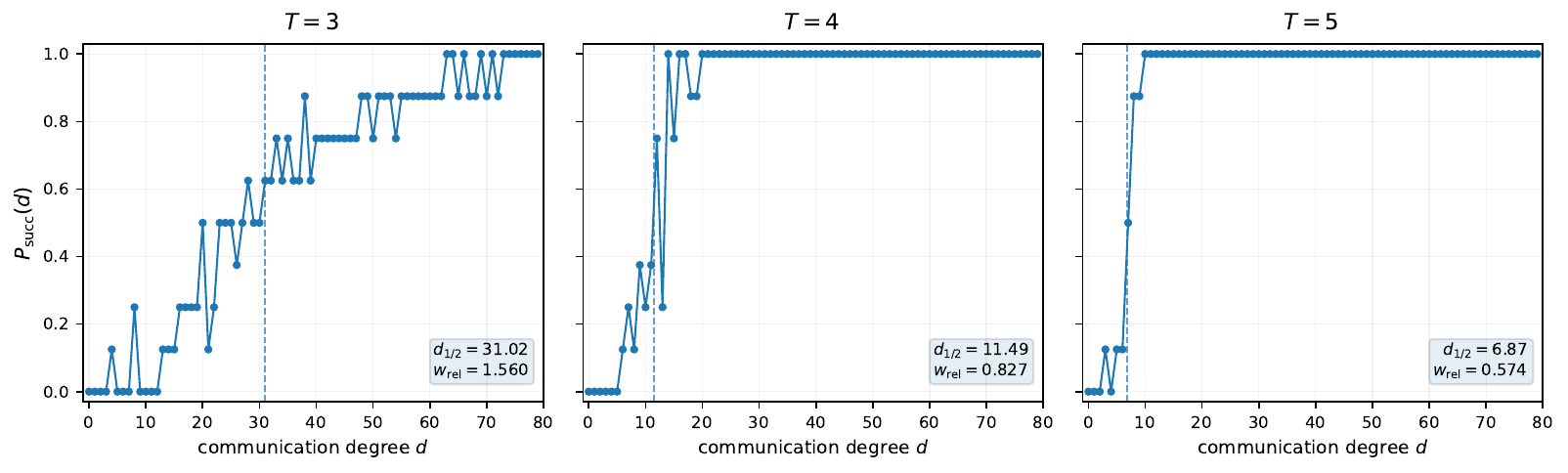}
    \caption{
    Empirical success probability $P_{\mathrm{succ}}(d)$ as a function of
communication degree $d$ for representative horizons $T=3,4,5$ in the
synthetic benchmark.
Each point is the observed success fraction over eight independent task seeds,
where an episode is successful if at least one agent recovers the ground-truth
hypothesis with supporting evidence.
Dashed vertical lines indicate the fitted crossover midpoint $d_{1/2}$.
As $T$ increases, the crossover shifts to lower communication degree and becomes
visibly sharper.
    }
    \label{fig:success_crossover_raw}
\end{figure}

Figure~\ref{fig:success_crossover_raw} shows the raw empirical
$P_{\mathrm{succ}}(d)$ for three representative horizons,
$T=3,4,5$. The crossover is broad at $T=3$ and becomes progressively
more concentrated in $d$ as $T$ increases.

To quantify this sharpening, we fit the episode-level binary success outcomes
with a logistic crossover,
\begin{equation}
P_{\mathrm{fit}}(d)
=
\frac{1}
{1+\exp[-(d-d_{1/2})/w]},
\label{eq:success_logistic}
\end{equation}
where $d_{1/2}$ is the degree at which the fitted success probability equals
$1/2$, and $w$ controls the crossover width. Let $d_p$ denote the degree at
which $P_{\mathrm{fit}}(d_p)=p$. We define the $10\%$--$90\%$ width as
\[
\Delta d_{10\text{--}90}
=
d_{0.9}-d_{0.1}
=
w\ln 81.
\]
Because the location of the crossover shifts substantially with $T$, we use
the dimensionless relative width
\begin{equation}
w_{\mathrm{rel}}
\equiv
\frac{\Delta d_{10\text{--}90}}{d_{1/2}}
=
\frac{w\ln 81}{d_{1/2}}.
\label{eq:relative_crossover_width}
\end{equation}
Smaller $w_{\mathrm{rel}}$ corresponds to a sharper crossover.

Across the full scan $T=3,\ldots,8$, we obtain
\[
\begin{array}{c|cccccc}
T
& 3 & 4 & 5 & 6 & 7 & 8 \\
\hline
w_{\mathrm{rel}}
& 1.560 & 0.827 & 0.574 & 0.443 & 0.436 & 0.069 .
\end{array}
\]
The relative width decreases strongly with $T$, with only a small
finite-sample variation between $T=6$ and $T=7$.

\begin{figure}[H]
    \centering
    \includegraphics[width=0.62\linewidth]{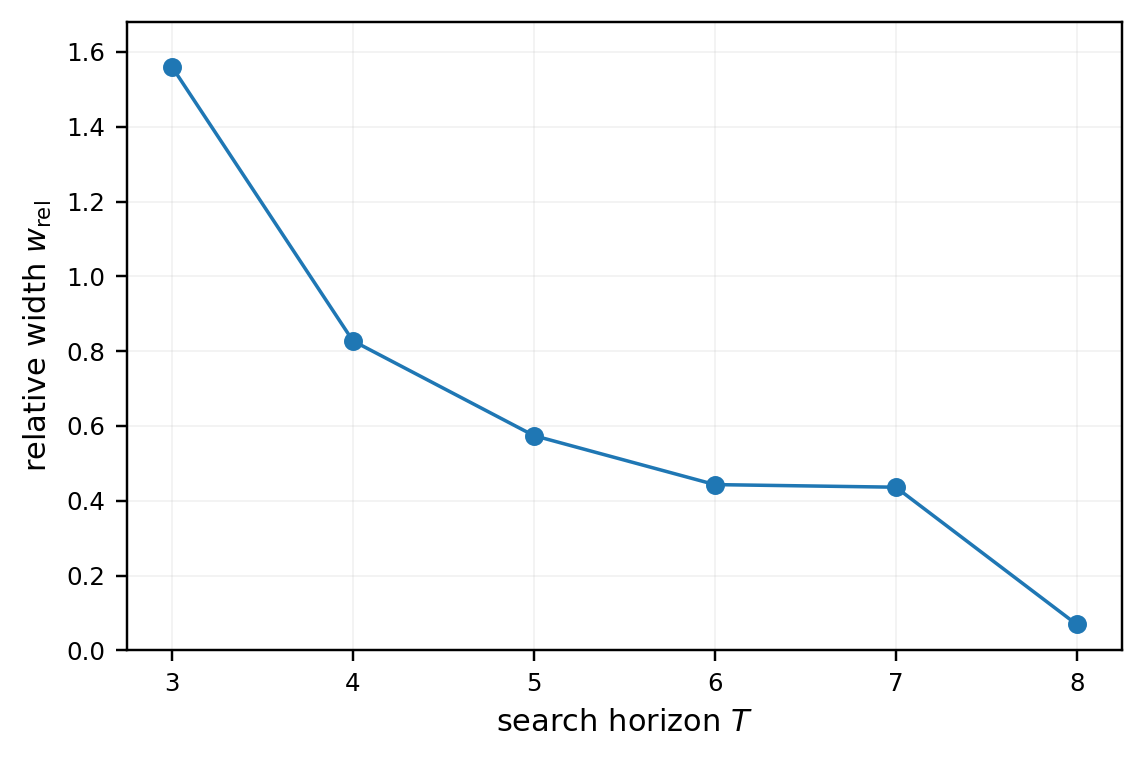}
    \caption{
    \textbf{Relative crossover width versus search horizon.}
    The relative $10\%$--$90\%$ width
    $w_{\mathrm{rel}}
    = (d_{0.9}-d_{0.1})/d_{1/2}$
    obtained from logistic fits to the episode-level success outcomes.
    The width decreases strongly as $T$ increases, indicating progressive
    sharpening of the finite-size success crossover.
    The line connecting the points is included only as a guide to the eye.
    }
    \label{fig:relative_width_vs_T}
\end{figure}

Together, Figs.~\ref{fig:success_crossover_raw}
and~\ref{fig:relative_width_vs_T} show that increasing the search horizon
systematically sharpens the finite-horizon success crossover in communication
degree. This progressive narrowing is consistent with increasingly
phase-transition-like behavior as $T$ grows, while the transition remains
smooth at the finite horizons considered here.

\subsection{LLM Communication Behavior under Free Communication}
\label{app:communication_behavior}

Table~\ref{tab:comm-honesty} shows a detailed breakdown of how agents chose to communicate under free communication policies Team-Free and Indiv-Free.
Both conditions allow free-form messages, and the prompt explicitly permits agents to withhold a result or to report an outcome different from the one they observed. A very small percentage of Qwen agents misreported, but the most noticeable result is that Qwen models are generally more reticent than GPT models. Qwen models communicate positive verification results 49-66\% of the time, while GPT models do 75-84\% of the time. Qwen models are also more unlikely to disclose negative verification results (35-53\%), compared to 69-84\% by GPT models.

\begin{table}
  \centering
    \caption{Communication behavior under Team-Free and Indiv-Free policies, pooled over six task seeds and three environments. }
\begingroup\setlength{\tabcolsep}{3pt}
\begin{tabular*}{\textwidth}{@{\extracolsep{\fill}}l*{8}{c}@{}}
\toprule
& \multicolumn{2}{c}{Qwen3.5-4B} & \multicolumn{2}{c}{Qwen3.5-9B} & \multicolumn{2}{c}{GPT-5.4 Nano} & \multicolumn{2}{c}{GPT-6 Luna} \\
\cmidrule(lr){2-3} \cmidrule(lr){4-5} \cmidrule(lr){6-7} \cmidrule(lr){8-9}
Metric & \makecell{Team-\\Free} & \makecell{Indiv-\\Free} & \makecell{Team-\\Free} & \makecell{Indiv-\\Free} & \makecell{Team-\\Free} & \makecell{Indiv-\\Free} & \makecell{Team-\\Free} & \makecell{Indiv-\\Free} \\
\midrule
Structured reports sent & 6{,}702 & 6{,}921 & 5{,}081 & 4{,}975 & 8{,}965 & 8{,}556 & 9{,}103 & 9{,}004 \\
Misreported outcomes & 0.81\% & 0.87\% & 0.98\% & 1.19\% & 0.09\% & 0.09\% & 0.00\% & 0.00\% \\
Sent nothing & 13.8\% & 15.2\% & 27.4\% & 30.5\% & 13.3\% & 17.7\% & 5.7\% & 16.3\% \\
Used free text & 68.8\% & 63.2\% & 68.7\% & 64.8\% & 75.5\% & 64.9\% & 89.9\% & 70.1\% \\
Positive findings disclosed & 66.1\% & 65.6\% & 49.8\% & 49.0\% & 80.3\% & 75.5\% & 84.0\% & 82.5\% \\
Negative findings disclosed & 51.5\% & 53.2\% & 36.1\% & 35.4\% & 72.5\% & 69.1\% & 84.3\% & 82.4\% \\
\bottomrule
\end{tabular*}
\endgroup

  \label{tab:comm-honesty}
\end{table}

Table~\ref{tbl:paired-diff} shows $d_c(\text{ColumnName}) - d_c(\text{Team-Auto})$: the difference between $d_c$ for each free-commmunication policy and the baseline, automatic communication.

\begin{figure}
    \centering
    \includegraphics[width=0.8\linewidth]{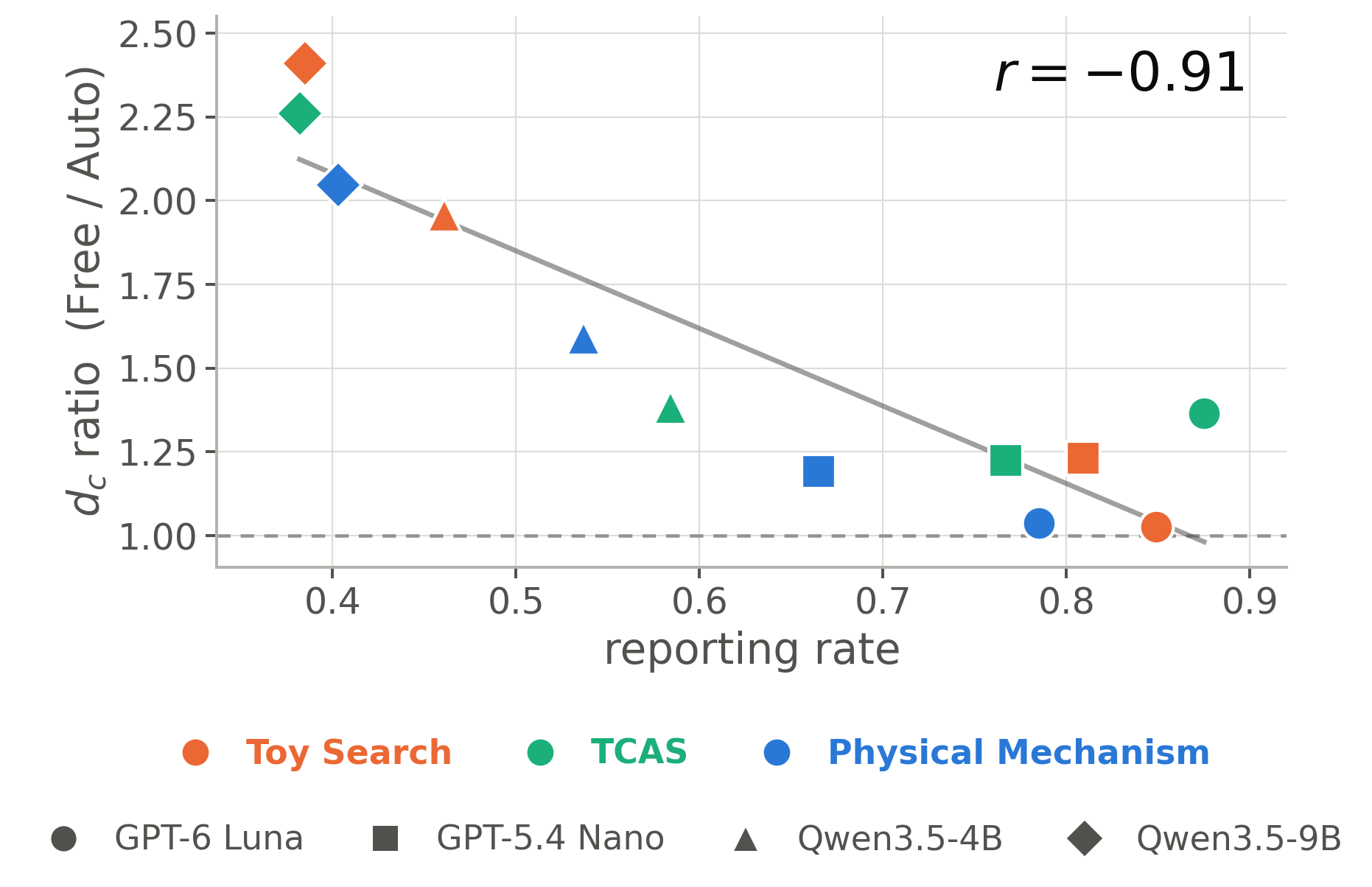}
    \caption{Correlation between ratio of Team-Free and Team-Auto $d_c$ values and the rate at which models report their findings to their neighbors.}
    \label{fig:reporting-rate}
\end{figure}

Figure~\ref{fig:reporting-rate} shows the correlation between the ratio of $d_c$ measured from Team-Free and Team-Auto policies (representing the difference in $d_c$ that perfect communication makes) and the rate at which models report their findings to their neighbors.

\begin{table}[]
    \centering
        \caption{ Each entry is
$\Delta d_c$, the paired difference in critical degree between a free-communication policy (Team-Free or Indiv-Free) and the fixed-communication baseline Team-Auto: positive means free communication \emph{raises} $d_c$, so more communication is needed to suppress incorrect lineages.
Intervals are exact 95\% cluster bootstraps over those six seeds. Entries are
$\Delta^{+\mathrm{upper}}_{-\mathrm{lower}}$.}
\begin{tabular*}{\textwidth}{@{\extracolsep{\fill}}l*{6}{c}@{}}
\toprule
& \multicolumn{2}{c}{Synthetic} & \multicolumn{2}{c}{TCAS} & \multicolumn{2}{c}{Physics} \\
\cmidrule(lr){2-3} \cmidrule(lr){4-5} \cmidrule(lr){6-7}
Model & \makecell{Team-\\Free} & \makecell{Indiv-\\Free} & \makecell{Team-\\Free} & \makecell{Indiv-\\Free} & \makecell{Team-\\Free} & \makecell{Indiv-\\Free} \\
\midrule
Qwen3.5-4B & \textbf{\ci{6.27}{0.48}{0.56}} & \textbf{\ci{6.92}{2.55}{1.90}} & \textbf{\ci{4.90}{6.77}{0.58}} & \textbf{\ci{7.33}{4.16}{2.66}} & \textbf{\ci{5.17}{1.09}{1.48}} & \textbf{\ci{5.54}{2.88}{3.10}} \\
Qwen3.5-9B & \textbf{\ci{9.37}{1.05}{2.03}} & \textbf{\ci{9.42}{2.00}{2.08}} & \textbf{\ci{16.24}{5.02}{5.47}} & \textbf{\ci{14.40}{1.20}{1.99}} & \textbf{\ci{8.73}{1.69}{2.18}} & \textbf{\ci{8.43}{1.72}{2.13}} \\
GPT-5.4 Nano & \textbf{\ci{1.48}{1.38}{0.84}} & \textbf{\ci{0.90}{0.66}{0.63}} & \textbf{\ci{2.35}{0.87}{1.62}} & \textbf{\ci{3.88}{0.82}{3.16}} & \textbf{\ci{2.09}{3.23}{1.97}} & \textbf{\ci{1.84}{0.86}{1.19}} \\
GPT-6 Luna & \ci{0.18}{0.60}{0.57} & \ci{-0.42}{0.67}{0.64} & \textbf{\ci{3.66}{3.51}{3.25}} & \ci{3.08}{0.96}{3.46} & \ci{0.32}{0.84}{0.97} & \ci{-0.28}{0.67}{0.54} \\
\bottomrule
\end{tabular*}

    \label{tbl:paired-diff}
\end{table}

% \begin{table}[]
%     \centering
%     \caption{impact of reasoning gpt-6-luna}
% % requires \usepackage{makecell} (and booktabs)
% \begin{tabular*}{\textwidth}{@{\extracolsep{\fill}}l*{9}{c}@{}}
% \toprule
% & \multicolumn{3}{c}{Synthetic} & \multicolumn{3}{c}{TCAS} & \multicolumn{3}{c}{Physics} \\
% \cmidrule(lr){2-4} \cmidrule(lr){5-7} \cmidrule(lr){8-10}
% \makecell[l]{Reasoning} & \makecell{Team-\\Auto} & \makecell{Team-\\Free} & \makecell{Indiv-\\Free} & \makecell{Team-\\Auto} & \makecell{Team-\\Free} & \makecell{Indiv-\\Free} & \makecell{Team-\\Auto} & \makecell{Team-\\Free} & \makecell{Indiv-\\Free} \\
% \midrule
% None & 6.88 & \textbf{7.06} & \textbf{6.46} & \textbf{10.05} & \textbf{13.71} & \textbf{13.13} & \textbf{8.17} & \textbf{8.48} &\textbf{ 7.88} \\
% Low & \textbf{6.78} & 7.11 & 6.85 & 15.64 & 37.59 & 29.16 & 8.45 & 8.84 & 8.76 \\
% Medium & 6.80 & 7.33 & 7.34 & 17.76 & 34.25 & n/a & 9.74 & 10.13 & 10.92 \\
% High & 6.98 & 7.02 & 7.07 & 15.30 & n/a & n/a & 12.18 & 17.20 & 16.58 \\
% XHigh & 7.08 & 7.23 & 7.57 & 18.82 & 26.97 & 28.98 & 19.64 & 22.01 & 27.70 \\
% \midrule
% \makecell[l]{Theory} & \multicolumn{3}{c}{6.10} & \multicolumn{3}{c}{10.90} & \multicolumn{3}{c}{6.10} \\
% \bottomrule
% \end{tabular*}
%     \label{tbl:reasoning}
% \end{table}

\subsection{Effect of Increasing Reasoning Effort}

\begin{figure}
    \centering
    \includegraphics[width=0.9\linewidth]{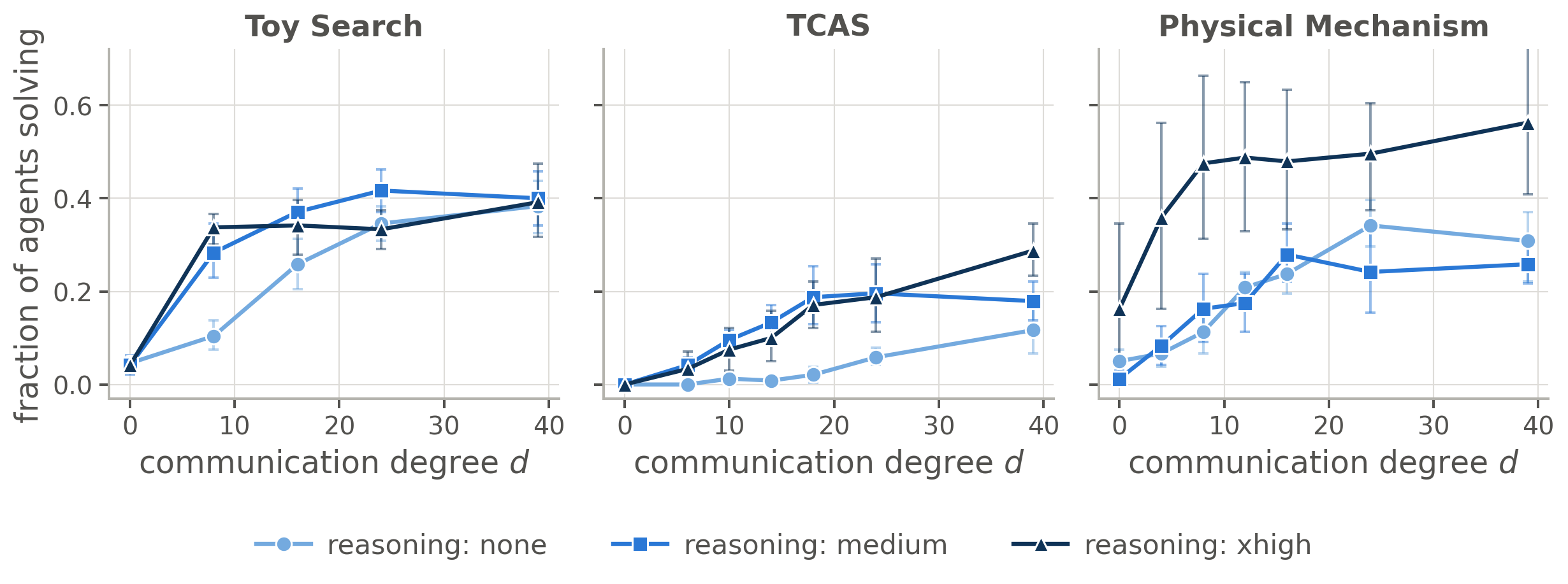}
    \caption{Fraction of agents completing the search increases with reasoning in GPT-6-luna.}
    \label{fig:phase-diag-reasoning}
\end{figure}

Figure~\ref{fig:phase-diag-reasoning} shows a nuance to the story that increased reasoning effort leads to worse results (higher $d_c$). 
While $d_c$ increases and deviates further from theory with more reasoning, reasoning often improves agents' ability to solve the task (especially on physical mechanism discovery task). This effect is true at higher communication degrees as well, suggesting that communication still helps.

\end{document}

%% file: math_commands.tex
\usepackage{amsmath,amsfonts,bm}

\newcommand{\cands}{\mathcal{C}}                    % candidate-component pool
\newcommand{\numcands}{\lvert\cands\rvert}          % number of candidates      (was M)
\newcommand{\numvals}{K}                            % values per component
\newcommand{\numagents}{N}                          % number of agents          (was N; kept)
\newcommand{\bagent}{b}                             % agent search branching    (was b)
\newcommand{\G}{\mathcal{G}}      % hypothesis graph (V,E)
\newcommand{\commdeg}{d}                            % communication degree
\newcommand{\critdeg}{d_{c}}                        % critical degree           (kept)
\newcommand{\numexp}{m}                             % experiments per round
\newcommand{\rounds}{T}                             % total rounds / GT depth
\newcommand{\Ainc}[1][t]{A_{#1}}                    % active incorrect hypotheses (was A_t)
\newcommand{\repfac}[1][t]{r_{#1}}                  % single-round reproduction factor
\newcommand{\pown}[1][t]{p^{\mathrm{own}}_{#1}}     % own-agent pruning
\newcommand{\pnbr}[1][t]{p^{\mathrm{nbr}}_{#1}}     % neighbor pruning
\def\eqref#1{equation~\ref{#1}}
\def\1{\bm{1}}

\DeclareMathAlphabet{\mathsfit}{\encodingdefault}{\sfdefault}{m}{sl}
\SetMathAlphabet{\mathsfit}{bold}{\encodingdefault}{\sfdefault}{bx}{n}

%% file: fig1.tex
\def\cmark{\ensuremath{\checkmark}}
\def\xmark{\ensuremath{\times}}
\providecommand{\Ainc}{A_{\mathrm{inc}}}
\providecommand{\commdeg}{d}
\providecommand{\critdeg}{d_c}
\providecommand{\rounds}{T}
\providecommand{\repfac}{R}
\providecommand{\pown}{p_{\mathrm{own}}}
\providecommand{\pnbr}{p_{\mathrm{nbr}}}
\providecommand{\bagent}{b}
\providecommand{\numagents}{N}

\colorlet{figcorr}{green!45!black}
\colorlet{figinc}{red!70!black}
\colorlet{figmsg}{blue!70!black}
\colorlet{figsel}{orange!85!black}

\tikzset{
  litbase/.style={rectangle, rounded corners=2pt, draw=black!60, fill=white,
                  minimum width=7mm, minimum height=4.8mm, inner sep=1pt, font=\scriptsize},
  parlit/.style={litbase, fill=black!8, draw=black!50},
  mk/.style={font=\tiny, inner sep=0.5pt},
  % states: u=unverified, p=pass, f=fail, x=pruned (own), r=pruned (received)
  L-u/.style={litbase},
  L-p/.style={litbase, draw=figcorr, fill=figcorr!20, thick,
              label={[mk, text=figcorr]below:\cmark}},
  L-f/.style={litbase, draw=figinc, fill=figinc!20, thick,
              label={[mk, text=figinc]below:\xmark}},
  L-x/.style={litbase, draw=black!30, dashed, fill=black!5, text=black!35,
              label={[mk, text=black!40]below:\xmark}},
  L-r/.style={litbase, draw=figmsg!70, dashed, fill=figmsg!5, text=figmsg!50,
              label={[mk, text=figmsg]below:\xmark}},
  E-u/.style={black!50}, E-p/.style={black!50}, E-f/.style={black!50},
  E-x/.style={black!20, dashed}, E-r/.style={figmsg!40, dashed},
}

% \minitree{prefix}{x}{state of l_{2,1}}{state of l_{2,2}}{state of l_{2,3}}
\newcommand{\minitree}[5]{%
  \begin{scope}[shift={(#2,0)}]
    \draw[black!30, densely dotted] (0,0) -- (0,0.5);
    \node[parlit] (#1P) at (0,0) {$\ell_{1,1}$};
    \node[L-#3] (#1A) at (-0.85,-1) {$\ell_{2,1}$};
    \node[L-#4] (#1B) at ( 0,   -1) {$\ell_{2,2}$};
    \node[L-#5] (#1C) at ( 0.85,-1) {$\ell_{2,3}$};
    \draw[E-#3] (#1P) -- (#1A);
    \draw[E-#4] (#1P) -- (#1B);
    \draw[E-#5] (#1P) -- (#1C);
  \end{scope}}

\begin{figure}[t]
\centering
\begin{tikzpicture}[>={Stealth[length=1.8mm]}]
  % step headers (equal spacing: 2.9)
  \foreach \x/\t in {0/{\textbf{1}~Select}, 2.9/{\textbf{2}~Verify},
                     5.8/{\textbf{3}~Receive}, 8.7/{\textbf{4}~Prune}}
    \node[font=\footnotesize, anchor=base] at (\x,0.8) {\t};
  % agent label
  % \node[rotate=90, font=\footnotesize] at (-2.0,-0.5) {agent $a_i$};
  % start candidate
  \node[parlit] (start) at (-3.1,0) {$\ell_{1,1}$};
  \node[font=\scriptsize, align=center, text=black!70, below=3pt of start]
       {candidate with a\\single hypothesis\\$\ell_{1,1}\equiv(q_1{=}1)$};
  \draw[->, black!35] (-1.95,-0.5) -- node[above, font=\tiny, text=black!60] {branch}
       node[below, font=\tiny, text=black!60] {$b{=}3$} (-1.3,-0.5);
  % states per step (l_{2,2} is on the correct lineage)
  \minitree{s1}{0}{u}{u}{u}
  \minitree{s2}{2.9}{f}{p}{u}
  \minitree{s4}{8.7}{x}{p}{r}
  % selection of m leaves
  \node[draw=figsel, thick, rounded corners=2pt, fit=(s1A)(s1B), inner sep=1.5pt,
        label={[font=\tiny, text=figsel, inner sep=1pt]below:$m{=}2$}] {};
  % flow arrows
  \draw[->, black!35] (1.28,-0.5) -- (1.62,-0.5);
  \draw[->, black!35] (4.18,-0.5) -- (7.42,-0.5);
  % incorrect active candidates and expected factors
  \foreach \x/\v in {0/2, 2.9/2}
    \node[font=\scriptsize] at (\x,-2.0) {incorrect active: \v};
  \node[font=\scriptsize, text=figcorr] (cnt) at (8.7,-2.0) {incorrect active: \textbf{0}};
  % \node[font=\scriptsize, text=black!60] at (8.7,-2.45)
  %      {$\times\,(1-\pown)(1-\pnbr)^{\commdeg}$};
  % external evidence (bottom aligned with the labels above)
  \coordinate (nbx) at (5.8,0);
  \node[draw=figmsg, fill=figmsg!5, rounded corners=3pt, font=\scriptsize,
        text=figmsg, align=center, anchor=south] (nb) at (cnt.south -| nbx)
       {from $\commdeg$ neighbors:\\$\{\ell_{1,1},\ell_{2,3}\}$\,\xmark};
  \draw[->, figmsg, thick] (nb.north) -- (5.8,-0.6);
  % \node[font=\scriptsize, text=black!60] at (8.7,-2.45)
  %      {$\times\,(1-\pown)(1-\pnbr)^{\commdeg}$};
  % next round
  % \draw[->, black!50, rounded corners=4pt] (8.7,-2.7) -- (8.7,-3.05)
  %      -- node[below, font=\scriptsize, text=black]
  %         {$\times\,\bagent$: each surviving candidate branches into $\bagent$ children (next round)}
  %      (0,-3.05) -- (0,-2.2);
  % legend
  % \foreach \s/\t/\x in {u/unverified/-0.9, p/pass/1.4, f/fail/3.0,
  %                       x/pruned (own)/4.6, r/pruned (received)/7.2} {
  %   \node[L-\s, minimum width=5mm] (lg\s) at (\x,-4.0) {};
  %   \node[font=\scriptsize, right=1pt of lg\s] {\t};
  % }
\end{tikzpicture}

\caption{\textbf{Schematic of our multi-agent search setup.} 
Agents begin with candidates each containing a single hypothesis, or \textit{literal} $\ell$ (in this example, the literal asserts the answer to question $q_1$ is 1). This candidate branches into $b$ new candidates that add each a new hypothesis. In one search round, an agent (1) selects $m$ of its active candidates to be verified; (2) runs the verification mechanism on the selected candidates to test whether they are consistent with the search target; (3) receives information about verification results from neighboring agents; and (4) the agent prunes incorrect candidates based on all available evidence.  The next round begins by branching $b$ new candidates from every surviving candidate of the current round.}
\label{fig:setup-round}
% \end{subfigure}
\end{figure}

%% file: taxonomy.tex
\definecolor{searchblue}{HTML}{2F64FF}
\definecolor{searchorange}{HTML}{D97A00}
\definecolor{searchgray}{HTML}{777777}
\definecolor{searchred}{HTML}{BB241A}

\tikzset{
  search node/.style={
    rectangle, rounded corners=1.5pt,
    draw=searchblue, fill=white, line width=0.9pt,
    minimum width=9mm, minimum height=6.5mm, inner sep=2.5pt},
  search truth/.style={draw=searchorange, line width=1.2pt},
  search refuted/.style={draw=searchgray, text=searchgray},
  search cut/.style={
    draw=searchred, dashed, line width=0.8pt,
    rounded corners=1.5pt, inner sep=2pt},
  search edge/.style={
    -{Latex[length=1.7mm,width=1.1mm]}, line width=0.65pt},
  target tag/.style={text=searchorange, font=\small, inner sep=1pt},
  qtag/.style={text=searchgray, font=\small, fill=white, inner sep=1.5pt},
  search panel/.style={x=1cm, y=1cm, font=\small}
}

\begin{figure}
\centering
\captionsetup[subfigure]{skip=5pt}
\tikzset{
  every node/.append style={font=\Large},            % plain annotation text
  search node/.append style={font=\Large},
  target tag/.append style={font=\Large},
  qtag/.append style={font=\Large, text=black!85},   % darker gray labels
}

% -------------------- (a) Needle --------------------
\begin{subfigure}[t]{0.48\linewidth}
\vspace{0pt}\centering
\resizebox{\linewidth}{!}{%
\begin{tikzpicture}[search panel]
  \path[use as bounding box] (-6.0,-2.2) rectangle (6.0,3.2);

  \node[search node,draw=black,line width=0.5pt]
    (pool) at (0,2.7) {literal pool $\mathcal{L}$};

  \node[search node]                (n1) at (-4.0,1.0) {$\ell_{1,1}$};
  \node[search node]                (n2) at (-2.0,1.0) {$\ell_{1,2}$};
  \node[search node,search refuted] (n3) at (0,1.0)    {$\ell_{1,3}$};
  \node[search node]                (n4) at (2.0,1.0)  {$\ell_{1,4}$};
  \node[search node,search truth]   (n5) at (4.0,1.0)  {$\ell_{1,5}$};

  \foreach \v in {n1,n2,n3,n4,n5}
    \draw[search edge] (pool) -- (\v);

  \node[search cut,fit=(n3)] {};
  \node[target tag] at (4.0,0.25) {$c^\star$};
  \node[text=searchred,align=center] at (0,-0.5)
    {refuting $\ell_{1,3}$ rules out\\one answer and nothing else};
\end{tikzpicture}}
\caption{\textbf{Needle.} All candidates have only one literal, so branching is impossible.}
\label{fig:needle-structure}
\end{subfigure}\hfill
% -------------------- (b) Chain --------------------
\begin{subfigure}[t]{0.48\linewidth}
\vspace{0pt}\centering
\resizebox{\linewidth}{!}{%
\begin{tikzpicture}[search panel]
  \path[use as bounding box] (-6.0,-2.2) rectangle (6.0,3.2);

  \node[search node]             (c0) at (-4.6,1.2) {$\varnothing$};
  \node[search node]             (c1) at (-2.6,1.2) {$\ell_{1,1}$};
  \node[search node]             (c2) at (0,1.2)    {$\ell_{1,1}\ell_{2,1}$};
  \node[search node,search truth](c3) at (3.3,1.2)
    {$\ell_{1,1}\ell_{2,1}\ell_{3,1}$};

  \foreach \a/\b in {c0/c1,c1/c2,c2/c3}
    \draw[search edge] (\a) -- (\b);

  \node[qtag] at (-3.6,1.75) {$q_1$};
  \node[qtag] at (-1.3,1.75) {$q_2$};
  \node[qtag] at (1.65,1.75) {$q_3$};

  \node[target tag] at (3.3,0.45) {$c^\star$};
  \node[align=center] at (0,-0.8)
    {$q_{i+1}$ is available only once $q_i$ is answered,\\
     and one answer is admissible at each step};
\end{tikzpicture}}
\caption{\textbf{Chain.} The questions are forced into a single order,
so there is one sequence of candidates.}
\label{fig:chain-structure}
\end{subfigure}

\par\vspace{1em}

% -------------------- (c) Tree / hierarchical --------------------
\begin{subfigure}[t]{0.48\linewidth}
\vspace{0pt}\centering
\resizebox{\linewidth}{!}{%
\begin{tikzpicture}[search panel]
  \path[use as bounding box] (-6.0,-2.2) rectangle (6.0,3.2);

  \node[search node] (r)  at (0,2.7) {$\varnothing$};
  \node[search node]                (n1) at (-3.7,1.3) {$\ell_{1,1}$};
  \node[search node]                (n2) at (0,1.3)    {$\ell_{1,2}$};
  \node[search node,search refuted] (n3) at (3.7,1.3)  {$\ell_{1,3}$};

  \node[search node]              (a1) at (-4.7,-0.6) {$\ell_{1,1}\ell_{2,1}$};
  \node[search node]              (a2) at (-2.7,-0.6) {$\ell_{1,1}\ell_{2,2}$};
  \node[search node,search truth] (b1) at (-1.0,-0.6) {$\ell_{1,2}\ell_{3,1}$};
  \node[search node]              (b2) at (1.0,-0.6)  {$\ell_{1,2}\ell_{3,2}$};
  \node[search node,search refuted](d1) at (2.7,-0.6) {$\ell_{1,3}\ell_{4,1}$};
  \node[search node,search refuted](d2) at (4.7,-0.6) {$\ell_{1,3}\ell_{4,2}$};

  \foreach \a/\b in {r/n1,r/n2,r/n3,
    n1/a1,n1/a2, n2/b1,n2/b2, n3/d1,n3/d2}
    \draw[search edge] (\a) -- (\b);

  \node[qtag] at (-3.7,0.35) {then asks $q_2$};
  \node[qtag] at (0,0.35)    {then asks $q_3$};
  \node[qtag] at (3.7,0.35)  {then asks $q_4$};
  \node[qtag] at (-5.05,1.3) {$q_1$};

  \node[target tag] at (-1.0,-1.25) {$c^\star$};
  \node[search cut,inner sep=5pt,fit=(n3)(d1)(d2)] {};
  \node[text=searchred,align=center] at (3.7,-1.75)
    {refuting $\ell_{1,3}$ kills one subtree};
\end{tikzpicture}}
\caption{\textbf{Tree / hierarchical.} The next question depends on
the previous answer, so e.g., $q_4$ is asked only under $\ell_{1,3}$.}
\label{fig:tree-structure}
\end{subfigure}\hfill
% -------------------- (d) Factorizable / confluent --------------------
\begin{subfigure}[t]{0.48\linewidth}
\vspace{0pt}\centering
\resizebox{\linewidth}{!}{%
\begin{tikzpicture}[search panel]
  \path[use as bounding box] (-6.0,-2.2) rectangle (6.0,3.2);

  \node[search node] (r)  at (0,2.7) {$\varnothing$};
  \node[search node]                (n1) at (-4.2,1.3) {$\ell_{1,1}$};
  \node[search node]                (n2) at (-1.4,1.3) {$\ell_{2,1}$};
  \node[search node,search refuted] (n3) at (1.4,1.3)  {$\ell_{3,1}$};
  \node[search node]                (n4) at (4.2,1.3)  {$\ell_{4,1}$};

  \node[search node]               (p12) at (-4.75,-0.6) {$\ell_{1,1}\ell_{2,1}$};
  \node[search node,search refuted](p13) at (-2.85,-0.6) {$\ell_{1,1}\ell_{3,1}$};
  \node[search node,search truth]  (p14) at (-0.95,-0.6) {$\ell_{1,1}\ell_{4,1}$};
  \node[search node,search refuted](p23) at (0.95,-0.6)  {$\ell_{2,1}\ell_{3,1}$};
  \node[search node]               (p24) at (2.85,-0.6)  {$\ell_{2,1}\ell_{4,1}$};
  \node[search node,search refuted](p34) at (4.75,-0.6)  {$\ell_{3,1}\ell_{4,1}$};

  \foreach \a/\b in {r/n1,r/n2,r/n3,r/n4}
    \draw[search edge] (\a) -- (\b);
  \foreach \a/\b in {
    n1/p12,n1/p13,n1/p14,
    n2/p12,n2/p23,n2/p24,
    n3/p13,n3/p23,n3/p34,
    n4/p14,n4/p24,n4/p34}
    \draw[search edge] (\a) -- (\b);

  \node[qtag] at (0,2.05) {any question, in any order};

  \node[target tag] at (-0.95,-1.25) {$c^\star$};
  \foreach \v in {n3,p13,p23,p34}
    \node[search cut,fit=(\v)] {};
  \node[text=searchred,align=center] at (1.9,-1.75)
    {refuting $\ell_{3,1}$ kills every candidate holding it};
\end{tikzpicture}}
\caption{\textbf{Factorizable.} Literals can combine in any order, leading to increasing overlap as the graph grows.}
\label{fig:graph-structure}
\end{subfigure}

\caption{\textbf{Our search task taxonomy.}
Orange marks the target $c^\star$, and red dashes mark what refuting a single literal can remove. 
The classes differ in how literals can be combined, and that difference determines evidence reusability and branching behavior.}
\label{fig:search-structures}
\end{figure}

%% file: addl_refs.bib
@misc{el2026physicsagentsstatisticalmechanics,
      title={Physics of Agents: Statistical Mechanics Predicts Collective Behavior of AI Agents}, 
      author={Batu El and Jinhee Paeng and Fatih Dinc and Shiye Su and Mete Erdogan and Aneesh Pappu and Haotian Ye and Wanjia Zhao and Surya Ganguli and James Zou},
      year={2026},
      eprint={2608.16578},
      archivePrefix={arXiv},
      primaryClass={cs.AI},
      url={https://arxiv.org/abs/2608.16578}, 
}

@article{song2026webswarm,
  title   = {{WebSwarm}: Recursive Multi-Agent Orchestration for Deep-and-Wide Web Search},
  author  = {Song, Xiaoshuai and Zhang, Liancheng and Zhao, Kangzhi and Zhu, Yutao and Wang, Zhongyuan and Dong, Guanting and Yang, Jinghan and Li, Han and Gai, Kun and Wen, Ji-Rong and Dou, Zhicheng},
  journal = {arXiv preprint arXiv:2607.08662},
  year    = {2026},
  doi     = {10.48550/arXiv.2607.08662},
  url     = {https://arxiv.org/abs/2607.08662}
}

@article{prabhakar2025enterprisedeepresearch,
  title   = {Enterprise Deep Research: Steerable Multi-Agent Deep Research for Enterprise Analytics},
  author  = {Prabhakar, Akshara and Ram, Roshan and Chen, Zixiang and Savarese, Silvio and Wang, Frank and Xiong, Caiming and Wang, Huan and Yao, Weiran},
  journal = {arXiv preprint arXiv:2510.17797},
  year    = {2025},
  doi     = {10.48550/arXiv.2510.17797},
  url     = {https://arxiv.org/abs/2510.17797}
}

@inproceedings{wong2026widesearch,
  title     = {{WideSearch}: Benchmarking Agentic Broad Info-Seeking},
  author    = {Wong, Ryan and Wang, Jiawei and Zhao, Junjie and Chen, Li and Gao, Yan and Zhang, Long and Zhou, Xuan and Wang, Zuo and Xiang, Kai and Zhang, Ge and Huang, Wenhao and Wang, Yang and Wang, Ke},
  booktitle = {International Conference on Learning Representations},
  pages     = {10012--10086},
  year      = {2026}
}

@misc{tang2025taskcomplexity,
  title         = {On the Importance of Task Complexity in Evaluating {LLM}-Based Multi-Agent Systems},
  author        = {Bohan Tang and Huidong Liang and Keyue Jiang and Xiaowen Dong},
  year          = {2025},
  eprint        = {2510.04311},
  archivePrefix = {arXiv},
  doi           = {10.48550/arXiv.2510.04311},
  url           = {https://arxiv.org/abs/2510.04311}
}

@misc{kim2025scaling,
  title         = {Towards a Science of Scaling Agent Systems},
  author        = {Yubin Kim and Ken Gu and Chanwoo Park and Chunjong Park and
                   Samuel Schmidgall and A. Ali Heydari and Yao Yan and Zhihan Zhang and
                   Yuchen Zhuang and Yun Liu and Mark Malhotra and Paul Pu Liang and
                   Hae Won Park and Yuzhe Yang and Xuhai Xu and Yilun Du and
                   Shwetak Patel and Tim Althoff and Daniel McDuff and Xin Liu},
  year          = {2025},
  eprint        = {2512.08296},
  archivePrefix = {arXiv},
  primaryClass  = {cs.AI},
  doi           = {10.48550/arXiv.2512.08296},
  url           = {https://arxiv.org/abs/2512.08296}
}

@misc{yang2025topological,
  title         = {Topological Structure Learning Should Be A Research Priority for {LLM}-Based Multi-Agent Systems},
  author        = {Jiaxi Yang and Mengqi Zhang and Yiqiao Jin and Hao Chen and Qingsong Wen and
                   Lu Lin and Yi He and Srijan Kumar and Weijie Xu and James Evans and Jindong Wang},
  year          = {2025},
  eprint        = {2505.22467},
  archivePrefix = {arXiv},
  primaryClass  = {cs.MA},
  doi           = {10.48550/arXiv.2505.22467},
  url           = {https://arxiv.org/abs/2505.22467}
}

@article{lai1984anomalies,
author = {Lai, Ten-Hwang and Sahni, Sartaj},
title = {Anomalies in parallel branch-and-bound algorithms},
year = {1984},
issue_date = {June 1984},
publisher = {Association for Computing Machinery},
address = {New York, NY, USA},
volume = {27},
number = {6},
issn = {0001-0782},
url = {https://doi.org/10.1145/358080.358103},
doi = {10.1145/358080.358103},
journal = {Commun. ACM},
month = jun,
pages = {594–602},
numpages = {9}
}

@INPROCEEDINGS{kuhn2006pseudo,
  author={Kuhn, D. Richard and Okun, Vadim},
  booktitle={2006 30th Annual IEEE/NASA Software Engineering Workshop}, 
  title={Pseudo-Exhaustive Testing for Software}, 
  year={2006},
  volume={},
  number={},
  pages={153-158},
  doi={10.1109/SEW.2006.26}}

@article{brunton2016sindy,
  author  = {Brunton, Steven L. and Proctor, Joshua L. and Kutz, J. Nathan},
  title   = {Discovering governing equations from data by sparse identification of nonlinear dynamical systems},
  journal = {Proceedings of the National Academy of Sciences},
  volume  = {113},
  number  = {15},
  pages   = {3932--3937},
  year    = {2016},
  doi     = {10.1073/pnas.1517384113}
}

@inproceedings{okawa2026biasedconsensus,
  title = {Emergence of Biased Consensus in Multi-Agent {LLM} Debates},
  author = {Okawa, Maya},
  booktitle = {Proceedings of the 43rd International Conference
               on Machine Learning},
  series = {Proceedings of Machine Learning Research},
  volume = {306},
  year = {2026},
  url = {https://arxiv.org/abs/2608.02827}
}

@misc{denobili2026collectivealignment,
  title = {Collective Alignment in {LLM} Multi-Agent Systems:
           Disentangling Bias from Cooperation via Statistical Physics},
  author = {De Nobili, Cristiano},
  year = {2026},
  eprint = {2605.10528},
  archivePrefix = {arXiv},
  url = {https://arxiv.org/abs/2605.10528}
}

@article{ashery2025conventions,
  title = {Emergent Social Conventions and Collective Bias
           in {LLM} Populations},
  author = {Ashery, Ariel Flint and Aiello, Luca Maria and
            Baronchelli, Andrea},
  journal = {Science Advances},
  volume = {11},
  number = {20},
  pages = {eadu9368},
  year = {2025},
  doi = {10.1126/sciadv.adu9368},
  url = {https://doi.org/10.1126/sciadv.adu9368}
}

@article{demarzo2026coordination,
  title = {{AI} Agents Can Coordinate Beyond Human Scale},
  author = {De Marzo, Giordano and Castellano, Claudio and
            Garcia, David},
  journal = {Science Advances},
  volume = {12},
  pages = {eaea6091},
  year = {2026},
  doi = {10.1126/sciadv.aea6091},
  url = {https://doi.org/10.1126/sciadv.aea6091}
}

@article{flint2026groupsize,
  title = {Group Size Effects and Collective Misalignment
           in {LLM} Multi-Agent Systems},
  author = {Flint, Ariel and Aiello, Luca Maria and
            Pastor-Satorras, Romualdo and Baronchelli, Andrea},
  journal = {Proceedings of the National Academy of Sciences},
  volume = {123},
  number = {34},
  pages = {e2531697123},
  year = {2026},
  doi = {10.1073/pnas.2531697123},
  url = {https://doi.org/10.1073/pnas.2531697123}
}

@article{morales2026nmr,
  title={NMR Elucidation as an Agentic Search Problem, Not a Modeling Problem},
  author={Morales, Irina Espejo and Hinz, Damon and Alberts, Marvin and Krawezik, Geraud and Jeong, Haewon and Ho, Shirley},
  journal={arXiv preprint arXiv:2607.19406},
  year={2026}
}

@inproceedings{chen2025mindsearch,
  title = {{MindSearch}: Mimicking Human Minds Elicits Deep {AI} Searcher},
  author = {Chen, Zehui and Liu, Kuikun and Wang, Qiuchen and
            Liu, Jiangning and Zhang, Wenwei and Chen, Kai and
            Zhao, Feng},
  booktitle = {International Conference on Learning Representations},
  year = {2025},
  url = {https://arxiv.org/abs/2407.20183}
}

@misc{jin2025hira,
  title = {{HiRA}: A Hierarchical Reasoning Framework for
           Decoupled Planning and Execution in Deep Search},
  author = {Jin, Jiajie and Li, Xiaoxi and Dong, Guanting and
            Zhang, Yuyao and Zhu, Yutao and Zhao, Yang and
            Qian, Hongjin and Dou, Zhicheng},
  year = {2025},
  eprint = {2507.02652},
  archivePrefix = {arXiv},
  url = {https://arxiv.org/abs/2507.02652}
}

@inproceedings{yang2026mosa,
  title = {Multi-{LLM} Collaborative Search for Complex Problem Solving},
  author = {Yang, Sen and Li, Yafu and Lam, Wai and Cheng, Yu},
  booktitle = {Findings of the Association for Computational
               Linguistics: ACL 2026},
  publisher = {Association for Computational Linguistics},
  pages = {42599--42614},
  year = {2026},
  doi = {10.18653/v1/2026.findings-acl.2115},
  url = {https://aclanthology.org/2026.findings-acl.2115/}
}

@inproceedings{li2026matsir,
  title = {{LLM} Inductive Reasoning Through Multi-Agent Enhanced
           {Monte Carlo} Tree Search},
  author = {Li, Xiang and Zhou, Yucheng and Wei, Xiangzhi and
            Shi, Zesheng and Wan, Haiyuan and Gong, Yifan and
            Liu, Fangming and Li, Jing},
  booktitle = {Findings of the Association for Computational
               Linguistics: ACL 2026},
  publisher = {Association for Computational Linguistics},
  pages = {23548--23562},
  year = {2026},
  doi = {10.18653/v1/2026.findings-acl.1178},
  url = {https://aclanthology.org/2026.findings-acl.1178/}
}

@inproceedings{qian2025macnet,
  title     = {Scaling Large Language Model-based Multi-Agent Collaboration},
  author    = {Qian, Chen and Xie, Zihao and Wang, Yifei and Liu, Wei
               and Zhu, Kunlun and Xia, Hanchen and Dang, Yufan
               and Du, Zhuoyun and Chen, Weize and Yang, Cheng
               and Liu, Zhiyuan and Sun, Maosong},
  booktitle = {International Conference on Learning Representations},
  year      = {2025},
  url       = {https://proceedings.iclr.cc/paper_files/paper/2025/hash/66a026c0d17040889b50f0dfa650e5e0-Abstract-Conference.html}
}

@inproceedings{zhou2026multi,
  title={Multi-agent design: Optimizing agents with better prompts and topologies},
  author={Zhou, Han and Wan, Xingchen and Sun, Ruoxi and Palangi, Hamid and Iqbal, Shariq and Vuli{\'c}, Ivan and Korhonen, Anna and Arik, Sercan},
  booktitle={International Conference on Learning Representations},
  volume={2026},
  pages={15844--15872},
  year={2026}
}

@inproceedings{shen-etal-2025-understanding,
    title = "Understanding the Information Propagation Effects of Communication Topologies in {LLM}-based Multi-Agent Systems",
    author = "Shen, Xu  and
      Liu, Yixin  and
      Dai, Yiwei  and
      Wang, Yili  and
      Miao, Rui  and
      Tan, Yue  and
      Pan, Shirui  and
      Wang, Xin",
    editor = "Christodoulopoulos, Christos  and
      Chakraborty, Tanmoy  and
      Rose, Carolyn  and
      Peng, Violet",
    booktitle = "Proceedings of the 2025 Conference on Empirical Methods in Natural Language Processing",
    month = nov,
    year = "2025",
    address = "Suzhou, China",
    publisher = "Association for Computational Linguistics",
    url = "https://aclanthology.org/2025.emnlp-main.623/",
    doi = "10.18653/v1/2025.emnlp-main.623",
    pages = "12347--12361",
    ISBN = "979-8-89176-332-6"
}

@inproceedings{jiang-etal-2026-dynamic,
    title = "Dynamic Generation of Multi {LLM} Agents Communication Topologies with Graph Diffusion Models",
    author = "Jiang, Eric Hanchen  and
      Li, Levina  and
      Wan, Frank  and
      Liang, Xiao  and
      Yin, Sophia  and
      Wu, Yuchen  and
      Li, Xinfeng  and
      Sun, Yizhou  and
      Wang, Wei  and
      Chang, Kai-Wei  and
      Wu, Ying Nian",
    editor = "Liakata, Maria  and
      Moreira, Viviane P.  and
      Zhang, Jiajun  and
      Jurgens, David",
    booktitle = "Proceedings of the 64th Annual Meeting of the {A}ssociation for {C}omputational {L}inguistics (Volume 1: Long Papers)",
    month = jul,
    year = "2026",
    address = "San Diego, California, United States",
    publisher = "Association for Computational Linguistics",
    url = "https://aclanthology.org/2026.acl-long.1764/",
    doi = "10.18653/v1/2026.acl-long.1764",
    pages = "38042--38060",
    ISBN = "979-8-89176-390-6"
}

@article{clearwater1991cooperative,
  title   = {Cooperative Solution of Constraint Satisfaction Problems},
  author  = {Clearwater, Scott H. and Huberman, Bernardo A.
             and Hogg, Tad},
  journal = {Science},
  volume  = {254},
  number  = {5035},
  pages   = {1181--1183},
  year    = {1991},
  doi     = {10.1126/science.254.5035.1181}
}

@article{hamadi2009manysat,
  title   = {{ManySAT}: A Parallel {SAT} Solver},
  author  = {Hamadi, Youssef and Jabbour, Said and Sais, Lakhdar},
  journal = {Journal on Satisfiability, Boolean Modeling and Computation},
  volume  = {6},
  pages   = {245--262},
  year    = {2009},
  url     = {https://www.cril.univ-artois.fr/~jabbour/PDF/jsatmanysat.pdf}
}

@article{dechter2007and,
  title   = {{AND/OR} Search Spaces for Graphical Models},
  author  = {Dechter, Rina and Mateescu, Robert},
  journal = {Artificial Intelligence},
  volume  = {171},
  number  = {2--3},
  pages   = {73--106},
  year    = {2007},
  doi     = {10.1016/j.artint.2006.11.003}
}

@inproceedings{dechter1986learning,
  title     = {Learning While Searching in Constraint-Satisfaction-Problems},
  author    = {Dechter, Rina},
  booktitle = {Proceedings of the Fifth National Conference
               on Artificial Intelligence},
  pages     = {178--183},
  year      = {1986},
  publisher = {AAAI Press},
  url       = {https://cdn.aaai.org/AAAI/1986/AAAI86-029.pdf}
}

@article{marinescu2009andor,
  title   = {{AND/OR} Branch-and-Bound Search for Combinatorial
             Optimization in Graphical Models},
  author  = {Marinescu, Radu and Dechter, Rina},
  journal = {Artificial Intelligence},
  volume  = {173},
  number  = {16--17},
  pages   = {1457--1491},
  year    = {2009},
  url     = {https://ics.uci.edu/~dechter/publications/r151.pdf}
}

@article{karp1993randomized,
  author  = {Karp, Richard M. and Zhang, Yanjun},
  title   = {Randomized Parallel Algorithms for Backtrack Search
             and Branch-and-Bound Computation},
  journal = {Journal of the ACM},
  volume  = {40},
  number  = {3},
  pages   = {765--789},
  year    = {1993}
}

@article{wu2023autogen,
  title={Autogen: Enabling next-gen llm applications via multi-agent conversation},
  author={Wu, Qingyun and Bansal, Gagan and Zhang, Jieyu and Wu, Yiran and Li, Beibin and Zhu, Erkang and Jiang, Li and Zhang, Xiaoyun and Zhang, Shaokun and Liu, Jiale and others},
  journal={arXiv preprint arXiv:2308.08155},
  year={2023}
}

@inproceedings{guo2024multiagents,
  title     = {Large Language Model Based Multi-Agents: A Survey of Progress and Challenges},
  author    = {Guo, Taicheng and Chen, Xiuying and Wang, Yaqi and Chang, Ruidi and Pei, Shichao and Chawla, Nitesh V. and Wiest, Olaf and Zhang, Xiangliang},
  booktitle = {Proceedings of the Thirty-Third International Joint Conference on Artificial Intelligence, {IJCAI}-24},
  pages     = {8048--8057},
  year      = {2024},
  month     = aug,
  publisher = {International Joint Conferences on Artificial Intelligence Organization},
  note      = {Survey Track},
  doi       = {10.24963/ijcai.2024/890},
  url       = {https://doi.org/10.24963/ijcai.2024/890}
}

@inproceedings{hong2024metagpt,
  title={MetaGPT: Meta programming for a multi-agent collaborative framework},
  author={Hong, Sirui and Zhuge, Mingchen and Chen, Jonathan and Zheng, Xiawu and Cheng, Yuheng and Wang, Jinlin and Zhang, Ceyao and Yau, Steven and Lin, Zijuan and Zhou, Liyang and others},
  booktitle={International Conference on Learning Representations},
  volume={2024},
  pages={23247--23275},
  year={2024}
}

@inproceedings{qian2024chatdev,
  title     = {{ChatDev}: Communicative Agents for Software Development},
  author    = {Qian, Chen and Liu, Wei and Liu, Hongzhang and Chen, Nuo and Dang, Yufan and Li, Jiahao and Yang, Cheng and Chen, Weize and Su, Yusheng and Cong, Xin and Xu, Juyuan and Li, Dahai and Liu, Zhiyuan and Sun, Maosong},
  booktitle = {Proceedings of the 62nd Annual Meeting of the Association for Computational Linguistics (Volume 1: Long Papers)},
  pages     = {15174--15186},
  year      = {2024},
  month     = aug,
  address   = {Bangkok, Thailand},
  publisher = {Association for Computational Linguistics},
  doi       = {10.18653/v1/2024.acl-long.810},
  url       = {https://aclanthology.org/2024.acl-long.810/}
}

@inproceedings{du2024multiagent,
  title     = {Improving Factuality and Reasoning in Language Models through Multiagent Debate},
  author    = {Du, Yilun and Li, Shuang and Torralba, Antonio and Tenenbaum, Joshua B. and Mordatch, Igor},
  booktitle = {Proceedings of the 41st International Conference on Machine Learning},
  series    = {Proceedings of Machine Learning Research},
  volume    = {235},
  pages     = {11733--11763},
  year      = {2024},
  publisher = {PMLR},
  url       = {https://proceedings.mlr.press/v235/du24e.html}
}

@inproceedings{kim2024mdagents,
  title     = {{MDAgents}: An Adaptive Collaboration of {LLM}s for Medical Decision-Making},
  author    = {Kim, Yubin and Park, Chanwoo and Jeong, Hyewon and Chan, Yik Siu and Xu, Xuhai and McDuff, Daniel and Lee, Hyeonhoon and Ghassemi, Marzyeh and Breazeal, Cynthia and Park, Hae Won},
  booktitle = {Advances in Neural Information Processing Systems},
  volume    = {37},
  pages     = {79410--79452},
  year      = {2024},
  doi       = {10.52202/079017-2522},
  url       = {https://proceedings.neurips.cc/paper_files/paper/2024/hash/90d1fc07f46e31387978b88e7e057a31-Abstract-Conference.html}
}

@article{Ising1925,
  author  = {Ising, Ernst},
  title   = {Beitrag zur Theorie des Ferromagnetismus},
  journal = {Zeitschrift f{\"u}r Physik},
  year    = {1925},
  volume  = {31},
  number  = {1},
  pages   = {253--258},
  doi     = {10.1007/BF02980577}
}

@inproceedings{li2024improving,
  title={Improving multi-agent debate with sparse communication topology},
  author={Li, Yunxuan and Du, Yibing and Zhang, Jiageng and Hou, Le and Grabowski, Peter and Li, Yeqing and Ie, Eugene},
  booktitle={Findings of the Association for Computational Linguistics: EMNLP 2024},
  pages={7281--7294},
  year={2024}
}

@book{gulwani2017program,
author = {Gulwani, Sumit and Polozov, Alex and Singh, Rishabh},
title = {Program Synthesis},
booktitle = {Foundations and Trends in Programming Languages},
year = {2017},
month = {August},
publisher = {NOW},
url = {https://www.microsoft.com/en-us/research/publication/program-synthesis/},
pages = {1-119},
volume = {4},
}

@article{sutton2019bitter,
  title={The bitter lesson},
  author={Sutton, Richard},
  year={2019}
}

@inproceedings{agashe2025llm,
  title={Llm-coordination: evaluating and analyzing multi-agent coordination abilities in large language models},
  author={Agashe, Saaket and Fan, Yue and Reyna, Anthony and Wang, Xin Eric},
  booktitle={Findings of the Association for Computational Linguistics: NAACL 2025},
  pages={8053--8072},
  year={2025}
}

@misc{tanaka2026collectiveintelligencelotterymultiagent,
      title={When Is Collective Intelligence a Lottery? Multi-Agent Scaling Laws for Memetic Drift in LLMs}, 
      author={Hidenori Tanaka},
      year={2026},
      eprint={2603.24676},
      archivePrefix={arXiv},
      primaryClass={cs.AI},
      url={https://arxiv.org/abs/2603.24676}, 
}

@misc{pavlova2026flaggametoymodel,
      title={Flag Game: A Toy Model for Mechanistic Swarm Interpretability}, 
      author={Elizabeth Pavlova and Hidenori Tanaka},
      year={2026},
      eprint={2609.19124},
      archivePrefix={arXiv},
      primaryClass={cs.AI},
      url={https://arxiv.org/abs/2609.19124}, 
}


%% file: references.bib
@inproceedings{ghandehari2013applying,
  author    = {Ghandehari, Laleh Shikh Gholamhossein and Borazjany, Mehra N.
               and Lei, Yu and Kacker, Raghu N. and Kuhn, D. Richard},
  title     = {Applying Combinatorial Testing to the {Siemens} Suite},
  booktitle = {2013 IEEE Sixth International Conference on Software Testing,
               Verification and Validation Workshops (ICSTW)},
  pages     = {362--371},
  year      = {2013},
  doi       = {10.1109/ICSTW.2013.47}
}

@inproceedings{hutchins1994experiments,
  author    = {Hutchins, Monica and Foster, Herb and Goradia, Tarak
               and Ostrand, Thomas},
  title     = {Experiments on the Effectiveness of Dataflow- and
               Controlflow-Based Test Adequacy Criteria},
  booktitle = {Proceedings of the 16th International Conference on
               Software Engineering (ICSE)},
  pages     = {191--200},
  year      = {1994}
}

@article{do2005supporting,
  author  = {Do, Hyunsook and Elbaum, Sebastian and Rothermel, Gregg},
  title   = {Supporting Controlled Experimentation with Testing Techniques:
             An Infrastructure and its Potential Impact},
  journal = {Empirical Software Engineering},
  volume  = {10},
  number  = {4},
  pages   = {405--435},
  year    = {2005},
  doi     = {10.1007/s10664-005-3861-2}
}

@article{zhuge2024gptswarm,
  title   = {{GPTSwarm}: Language Agents as Optimizable Graphs},
  author  = {Zhuge, Mingchen and Wang, Wenyi and Kirsch, Louis and Faccio, Francesco and Khizbullin, Dmitrii and Schmidhuber, J{\"u}rgen},
  journal = {arXiv preprint arXiv:2402.16823},
  year    = {2024},
  eprint  = {2402.16823},
  archivePrefix = {arXiv},
  primaryClass  = {cs.AI}
}

@article{zhang2024gdesigner,
  title   = {{G-Designer}: Architecting Multi-Agent Communication Topologies via Graph Neural Networks},
  author  = {Zhang, Guibin and Yue, Yanwei and Sun, Xiangguo and Wan, Guancheng and Yu, Miao and Fang, Junfeng and Wang, Kun and Chen, Tianlong and Cheng, Dawei},
  journal = {arXiv preprint arXiv:2410.11782},
  year    = {2024},
  eprint  = {2410.11782},
  archivePrefix = {arXiv},
  primaryClass  = {cs.MA}
}

@article{xie2026errorcascades,
  title   = {From Spark to Fire: Modeling and Mitigating Error Cascades in LLM-Based Multi-Agent Collaboration},
  author  = {Xie, Yizhe and Zhu, Congcong and Zhang, Xinyue and Zhu, Tianqing and Ye, Dayong and Qi, Minfeng and Chen, Huajie and Zhou, Wanlei},
  journal = {arXiv preprint arXiv:2603.04474},
  year    = {2026},
  eprint  = {2603.04474},
  archivePrefix = {arXiv}
}

@article{zhu2025multiagentbench,
  title   = {{MultiAgentBench}: Evaluating the Collaboration and Competition of {LLM} Agents},
  author  = {Zhu, Kunlun and Du, Hongyi and Hong, Zhaochen and Yang, Xiaocheng and Guo, Shuyi and Wang, Zhe and Wang, Zhenhailong and Qian, Cheng and Tang, Xiangru and Ji, Heng and You, Jiaxuan},
  journal = {arXiv preprint arXiv:2503.01935},
  year    = {2025},
  eprint  = {2503.01935},
  archivePrefix = {arXiv},
  primaryClass  = {cs.MA}
}

@inproceedings{zhou2026mass,
  title     = {Multi-Agent Design: Optimizing Agents with Better Prompts and Topologies},
  author    = {Zhou, Han and Wan, Xingchen and Sun, Ruoxi and Palangi, Hamid and Iqbal, Shariq and Vuli{\'c}, Ivan and Korhonen, Anna and Ar{\i}k, Sercan {\"O}.},
  booktitle = {International Conference on Learning Representations},
  year      = {2026},
  note      = {MASS; arXiv:2502.02533}
}

@article{rizvimartel2025benefits,
  title   = {Benefits and Limitations of Communication in Multi-Agent Reasoning},
  author  = {Rizvi-Martel, Michael and Bhattamishra, Satwik and Rathi, Neil and Rabusseau, Guillaume and Hahn, Michael},
  journal = {arXiv preprint arXiv:2510.13903},
  year    = {2025},
  eprint  = {2510.13903},
  archivePrefix = {arXiv},
  primaryClass  = {cs.MA}
}

@article{hinrichsen2000absorbing,
  title   = {Non-equilibrium Critical Phenomena and Phase Transitions into Absorbing States},
  author  = {Hinrichsen, Haye},
  journal = {Advances in Physics},
  volume  = {49},
  number  = {7},
  pages   = {815--958},
  year    = {2000},
  doi     = {10.1080/00018730050198152}
}

@article{semerjian2003randomwalksat,
  title   = {Relaxation and Metastability in a Local Search Procedure for the Random Satisfiability Problem},
  author  = {Semerjian, Guilhem and Monasson, R{\'e}mi},
  journal = {Physical Review E},
  volume  = {67},
  number  = {6},
  pages   = {066103},
  year    = {2003},
  doi     = {10.1103/PhysRevE.67.066103}
}

@article{lee2010finitesize,
  title   = {Finite-Size Scaling in Random {$K$}-Satisfiability Problems},
  author  = {Lee, Sang Hoon and Ha, Meesoon and Jeon, Chanil and Jeong, Hawoong},
  journal = {Physical Review E},
  volume  = {82},
  number  = {6},
  pages   = {061109},
  year    = {2010},
  doi     = {10.1103/PhysRevE.82.061109}
}

@article{niu2026reliability,
  title         = {Reliability-Contagion Feasibility in {LLM} Multi-Agent Networks},
  author        = {Niu, Ruiwu and Shu, Xincheng and Zhao, Ying},
  journal       = {arXiv preprint arXiv:2607.21912},
  year          = {2026},
  eprint        = {2607.21912},
  archivePrefix = {arXiv}
}

@article{jin2026hypothesistree,
  title   = {Toward Generalist Autonomous Research via Hypothesis-Tree Refinement},
  author  = {Jin, Jiajie and Hu, Yuyang and Qiu, Kai and Dai, Qi and Luo, Chong and Dong, Guanting and Li, Xiaoxi and Zhao, Tong and Ma, Xiaolong and Zhang, Gongrui and Wu, Zhirong and Liu, Bei and Yang, Zhengyuan and Li, Linjie and Wang, Lijuan and Qian, Hongjin and Zhu, Yutao and Dou, Zhicheng},
  journal = {arXiv preprint arXiv:2606.11926},
  year    = {2026},
  eprint  = {2606.11926},
  archivePrefix = {arXiv}
}

@inproceedings{dascoli2024odeformer,
  title     = {ODEFormer: Symbolic Regression of Dynamical Systems with Transformers},
  author    = {d'Ascoli, St{\'e}phane and Becker, S{\"o}ren and
               Schwaller, Philippe and Mathis, Alexander and
               Kilbertus, Niki},
  booktitle = {The Twelfth International Conference on Learning Representations},
  year      = {2024},
  url       = {https://arxiv.org/abs/2310.05573}
}

@inproceedings{zheng2026newtonbench,
  title     = {NewtonBench: Benchmarking Generalizable Scientific Law Discovery in LLM Agents},
  author    = {Zheng, Tianshi and Tam, Kelvin Kiu-Wai and
               Nguyen, Newt Hue-Nam K. and Xu, Baixuan and
               Wang, Zhaowei and Cheng, Jiayang and
               Tsang, Hong Ting and Wang, Weiqi and
               Bai, Jiaxin and Fang, Tianqing and
               Song, Yangqiu and Wong, Ginny Y. and See, Simon},
  booktitle = {International Conference on Learning Representations},
  year      = {2026},
  url       = {https://arxiv.org/abs/2510.07172}
}
